\documentclass[twocolumn]{aastex62}
\usepackage{amssymb, amsmath}
\usepackage{graphicx}
\usepackage{natbib}
\usepackage{latexsym}
\usepackage{mathrsfs}
\usepackage{mathtools}
\usepackage{multirow}
\usepackage{color, colortbl}
\usepackage{booktabs,makecell}
\usepackage{rotating}
\usepackage{threeparttable}

\begin{document}

\title{Simultaneous Optical Transmission Spectroscopy of a Terrestrial, Habitable-zone Exoplanet\\with Two Ground-Based Multiobject Spectrographs}

\correspondingauthor{Hannah Diamond-Lowe}
\email{hdiamondlowe@cfa.harvard.edu}

\author[0000-0001-8274-6639]{Hannah Diamond-Lowe}
\affil{Center for Astrophysics $\vert$ Harvard \& Smithsonian, 60 Garden St., Cambridge, MA 02138, USA}

\author[0000-0002-3321-4924]{Zachory Berta-Thompson}
\affil{Department of Astrophysical and Planetary Sciences, University of Colorado, 2000 Colorado Ave., Boulder, CO 80305, USA}

\author[0000-0002-9003-484X]{David Charbonneau}
\affiliation{Center for Astrophysics $\vert$ Harvard \& Smithsonian, 60 Garden St., Cambridge, MA 02138, USA}

\author[0000-0001-7730-2240]{Jason Dittmann}
\affiliation{Department of Earth, Atmospheric and Planetary Sciences, Massachusetts Institute of Technology, 77 Massachusetts Ave., 54-918, Cambridge, MA 02139, USA}

\author[0000-0002-1337-9051]{Eliza M.-R. Kempton}
\affiliation{Department of Astronomy, University of Maryland, College Park, MD 20742, USA}

\received{August 24, 2019}
%\revised{April 6, 2020}
\revised{May 12, 2020}
\accepted{May 15, 2020}
\submitjournal{\aj}
 
\begin{abstract}
Investigating the atmospheres of rocky exoplanets is key to performing comparative planetology between these worlds and the terrestrial planets that reside in the inner solar system. Terrestrial exoplanet atmospheres exhibit weak signals, and attempting to detect them pushes at the boundaries of what is possible for current instrumentation. We focus on the habitable-zone terrestrial exoplanet LHS 1140b. Given its 25-day orbital period and 2 hr transit duration, capturing transits of LHS 1140b is challenging. We observed two transits of this object, approximately 1 yr apart, which yielded four data sets thanks to our simultaneous use of the IMACS and LDSS3C multiobject spectrographs mounted on the twin Magellan telescopes at Las Campanas Observatory. We present a jointly fit white light curve, as well as jointly fit 20 nm wavelength-binned light curves from which we construct a transmission spectrum. Binning the joint white light-curve residuals to 3-minute time bins gives an rms of 145 ppm; binning down to 10-minute time bins gives an rms of 77 ppm. Our median uncertainty in $R_p^2/R_s^2$ in the 20 nm wavelength bins is 260 ppm, and we achieve an average precision of 1.3$\times$ the photon noise when fitting the wavelength-binned light curves with a Gaussian process regression. Our precision on $R_p^2/R_s^2$ is a factor of four larger than the feature amplitudes of a clear, hydrogen-dominated atmosphere, meaning that we are not able to test realistic models of LHS 1140b's atmosphere. The techniques and caveats presented here are applicable to the growing sample of terrestrial worlds in the Transiting Exoplanet Survey Satellite era, as well as to the upcoming generation of ground-based giant segmented mirror telescopes.
\end{abstract}


\section{Introduction} \label{sec:intro}

Planetary atmospheres hold clues about surface processes, formation histories, and the potential for habitability for the planets they surround. Under the right circumstances, they can also reveal the presence of life on other worlds via biomarker gases \citep[e.g.,][and references therein]{Domagal-Goldman2011,Meadows2017}. In the solar system we see a great diversity of atmospheres, from the puffy hydrogen and helium envelopes around Jupiter and Saturn to the heavy carbon dioxide layer around Venus and the nitrogen-rich sky of Titan. 

The terrestrial bodies of the solar system boast a wide variety of atmospheric compositions and masses, but all are secondary, high mean molecular weight atmospheres. Results from the Kepler mission, combined with statistical and empirical follow-up, reveal that such worlds also exist in abundance outside the solar system, along with a completely new kind of terrestrial planet that has retained a hydrogen- and helium-dominated envelope \citep{Fressin2013}. For planets with radii $<10 R_{\oplus}$, those with radii $>1.6 R_{\oplus}$ have low bulk densities and likely host puffy hydrogen and helium envelopes captured from the stellar nebula, while those with radii $<1.6 R_{\oplus}$ are rocky in nature and likely host high mean molecular weight secondary atmospheres \citep{Owen&Wu2013,Lopez2013,Rogers2015,Fulton2017,Fulton2018,VanEylen2018}, though given the difficulties in detecting secondary atmospheres around small planets, we have not yet spectroscopically characterized any. The $1.6 R_{\oplus}$ mark is not a hard cutoff. Another way to look at this is that planets with bulk densities less than that of rock ($2.5-3.0$ g/cm$^3$) must have significantly large envelopes of hydrogen and helium in order to explain their low masses relative to their radii, whereas planets with bulk densities at or above that of rock are likely compositionally similar to the terrestrial objects found in the solar system.

To understand the rocky exoplanets, we must probe their atmospheres and determine their compositions. In this paper we focus on the technique of transmission spectroscopy, whereby observations of a planet's transit across its star, taken over a range of wavelengths, can reveal the planet's atmospheric composition, since different molecules absorb stellar light at different wavelengths. Within the limits of current instrumentation, we begin the exploration of small-planet atmospheres by looking for small planets that orbit the small stars closest to us. This is a simple function of the planet-to-star radius ratio $R_p/R_s$ (the larger the ratio, the easier it is to detect the planet) and the need for high signal-to-noise ratio measurements to differentiate the planet radius at one wavelength from another (the closer the star, the more photons can be collected per observation).

Before the launch of the Transiting Exoplanet Survey Satellite \citep[TESS;][]{Ricker2015}, the ground-based transit surveys MEarth and TRAPPIST \citep{Nutzman2008,Gillon2013,Irwin2015} discovered a handful of small planets around three small, nearby stars: GJ 1132, TRAPPIST-1, and LHS 1140, which follow-up observations by the Spitzer Space Telescope and K2 confirmed and, for the TRAPPIST-1 and LHS 1140 systems, bolstered with additional planet discoveries \citep{Berta-Thompson2015,Gillon2017a,Dittmann2017a,Dittmann2017b,Ment2019}. Follow-up by radial velocity instruments such as the High Accuracy Radial-velocity Planet Searcher \citep[HARPS;][]{Pepe2004} provided masses for planets in the GJ 1132 and LHS 1140 systems, thereby confirming their rocky natures. HARPS also discovered an additional, nontransiting planet in the GJ 1132 system \citep{Bonfils2018}. In the case of the two nearest terrestrial planets HD 219134b,c \citep{Gillon2017b}, their presence was detected via radial velocities from HARPS-North \citep{Cosentino2012} and were later found to transit by Spitzer. The dimness of TRAPPIST-1 makes radial velocity measurements challenging, so masses for the TRAPPIST-1 planets are instead estimated using transit timing variations \citep[TTVs;][]{Grimm2018}, revealing that some of the TRAPPIST-1 planets may have bulk densities comparable to that of water. Now in the era of TESS the sample of terrestrial exoplanets orbiting small ($<0.3\ R_{\odot}$), nearby ($<15$ pc) stars is growing, with LHS 3844b and LTT 1445Ab added recently \citep{Vanderspek2019,Winters2019}.

Though the presence of these terrestrial planets provides a tantalizing opportunity for atmospheric follow-up, the most we are able to do with current instrumentation is rule out the lowest mean molecular weight atmospheres dominated by hydrogen and helium, which confirms the aforementioned work on Kepler planets with radii $<10\ R_{\oplus}$. So far cloud-free low mean molecular weight atmospheres are ruled out for TRAPPIST-1b,c,d,e,f and for GJ 1132b \citep{deWit2016,deWit2018,Diamond-Lowe2018}.

With the goal of eventually detecting atmospheric biomarkers on habitable-zone worlds, we designed a project to characterize the atmosphere of LHS 1140b \citep{Dittmann2017a}, a habitable-zone terrestrial exoplanet orbiting a nearby mid-M dwarf. Since the planet's discovery, and our subsequent observing program, Data Release 2 of the Gaia mission \citep{GaiaMission2016,GaiaDR22018} moved LHS 1140 farther away than was initially thought, to its current distance of $14.993 \pm 0.015$ pc. This means that the stellar radius of LHS 1140 is larger than initially thought, which in turn increases the derived planet radius. With this new information, we find that LHS 1140b has a radius of $1.727 \pm 0.032\ R_{\oplus}$ and a mass of $6.98 \pm 0.89\ M_{\oplus}$, making its density of $7.5 \pm 1.0$ g/cm$^3$ consistent with a terrestrial composition \citep{Ment2019}. The planet's surface gravity is $23.7 \pm 2.7$ m/s$^2$ with an estimated effective temperature $T_{\mathrm{eff}} = 235 \pm 5$ K, assuming an albedo of zero. The atmospheric scale height of a planet is directly proportional to the planet's temperature and inversely proportional to its surface gravity. In the case of LHS 1140b, its atmospheric scale height and therefore the amplitudes of its atmospheric features are below what is detectable with our observations.

We note that when we began this project we assumed a lower surface gravity for the planet. This came about because the initial mass and radius estimates of LHS 1140b gave a bulk density consistent with a composition of more than $50$\% iron, which is implausible and in stark defiance of conventional planetary formation scenarios \citep{Zeng2016,Dittmann2017a}. As such, it seemed likely that the mass of LHS 1140b would be refined and lowered in a subsequent season of radial velocity measurements (Figure 2 of \citet{Morley2017} provides an illustration of this thinking). At the start of this project we adopted values for the stellar distance and planet radius from \citet{Dittmann2017a}. In predicting the atmospheric signal we assumed a terrestrial core-mass fraction, which implied a surface gravity of 17.5 m/s$^2$. With Gaia DR2 it became apparent that the initial measured mass of LHS 1140b was actually correct, and it was the initial radius measurement that was wrong. The revised parameters imply a terrestrial composition and yield a surface gravity of 23.7 m/s$^2$. While many parallaxes of nearby stars were refined by the Gaia mission, LHS 1140 was a particularly pathological case owing to how sparse its field is. For a relatively bright star like LHS 1140, pre-Gaia parallaxes are generally reliable, but if there are too few additional stars in the observing field against which to compare the position, the derived distance is unreliable. 

Despite the difficulty involved in detecting the atmosphere of LHS 1140b, it is one of the few terrestrial planets orbiting a nearby M star for which liquid water could potentially exist on the planet surface. However, equilibrium temperature is not the sole determinant for habitability. M stars like LHS 1140 spend more time in the pre-main-sequence phase than G stars like the Sun before settling onto the main-sequence branch \citep{Baraffe2002,Baraffe2015}. This means that M stars have longer periods of high-energy activity that can strip the atmospheres of the planets orbiting them \citep{Luger2015}. However, some high-energy flux, particularly in the near-ultraviolet (NUV), may be necessary to jump-start life \citep{Ranjan2017}.

\citet{Spinelli2019} use the UV and X-ray capabilities of the space-based \textit{Swift} observatory to investigate the high-energy nature of LHS 1140. They find that while LHS 1140 exhibits low levels of UV activity, its relatively high ratio of far-ultraviolet (FUV) to NUV flux could produce O$_2$ and H$_2$O abiotically through the dissociation of CO$_2$. The low amounts of NUV received by LHS 1140b (2\% of the amount that Earth receives) may not provide enough of a spark for abiogenesis. However, these current measurements of LHS 1140 do not represent its past levels of UV radiation. Detecting the atmosphere around LHS 1140b would provide a clue to past behavior of LHS 1140, and vice versa.

In this work, we are not ultimately able to investigate the atmosphere of LHS 1140b, illustrating the need for more observationally accessible habitable-zone terrestrial planet targets. LHS 1140b has an orbital period of $24.736959 \pm 0.000080$ days and a transit duration of 2.1 hr \citep{Ment2019}, making transits of this object rare and difficult to observe owing to the 6 hr of observing time necessary to capture both the transit and adequate baseline on either side from which to measure the depth. Spitzer observed transits of LHS 1140b in its 4.5 $\mu$m broadband photometric bandpass \citep[DDT Program 13174, PI Dittmann;][]{Ment2019}; this infrared point complements the optical observations we undertake here.

In this paper we present our observing program in Section~\ref{sec:obs}. We detail our data extraction process, along with an illustrative diagram, in Section~\ref{sec:extract}. We then detail the analysis of our extracted spectra in Section~\ref{sec:analysis}. The results of this work, along with a discussion of their implications, are presented in Section~\ref{sec:results}, followed by our conclusions in Section~\ref{sec:conclusion}.

\section{Observations} \label{sec:obs}

Given the period (24.7 days) and transit duration (2.1 hr) of LHS 1140b, opportunities to observe a complete transit of this object from Las Campanas Observatory in Chile, where we could also capture data before and after transit, were rare. However, the 2 hr transit duration offers the advantage that a single transit observation yields a high signal-to-noise ratio measurement of the transit depth. In 2017 and 2018, there was one opportunity per year to observe a complete transit of LHS 1140b, along with baseline before and after transit.

We were awarded two nights on the Magellan I (Baade) and Magellan II (Clay) telescopes through the Center for Astrophysics $\vert$ Harvard \& Smithsonian (PI Diamond-Lowe) to simultaneously observe the 2017 and 2018 transits of LHS 1140b with both telescopes. We used the IMACS \citep{Dressler2011} and LDSS3C \citep{Stevenson2016a} multiobject spectrographs on Baade and Clay, respectively, to observe the transit across the optical and near infrared spectrum. We were able to capture both transits, yielding a total of four data sets (two with IMACS and two with LDSS3C) for our project. The details of these observations are presented in Table~\ref{tab:obs}.

\begin{table*}[ht]
\centering
\caption{Observations with Magellan I (Baade) and Magellan II (Clay)\label{tab:obs}}
\begin{tabular}{ccccccC}
\tablewidth{0pt}
\hline
\hline
Data Set & Date & Exposure Time & Duty Cycle & Number of & Minimum & \mathrm{Seeing}\\
(Instrument, Year)      & (UTC) & (s)           & (\%) & Exposures & Air Mass        & \mathrm{(arcsec)} \\
\hline
IMACS 2017*  & 2017 Oct 27, 00:37:10 -- 07:09:02 & 15 & 32.3 & 510 & 1.029 & 0.60 \\
LDSS3C 2017 & 2017 Oct 27, 00:28:15 -- 07:14:14 & 15 & 46.9 & 766 & 1.029 & 0.80 \\
IMACS 2018  & 2018 Nov 02, 00:34:47 -- 07:15:41 & 15 & 32.3 & 508 & 1.029 & 0.50 \\
LDSS3C 2018 & 2018 Nov 02, 01:10:51 -- 07:11:45 & 15 & 46.9 & 686 & 1.029 & 0.40 \\
\hline
\end{tabular}
\begin{minipage}[t]{0.95\linewidth}
\hfill\break
{* Due to instrument systematics discussed in Section~\ref{subsec:IMACSextract}, we do no include this data set in the analysis.}
\end{minipage}
\end{table*}

\begin{figure}[!ht]
\includegraphics[width=0.465\textwidth]{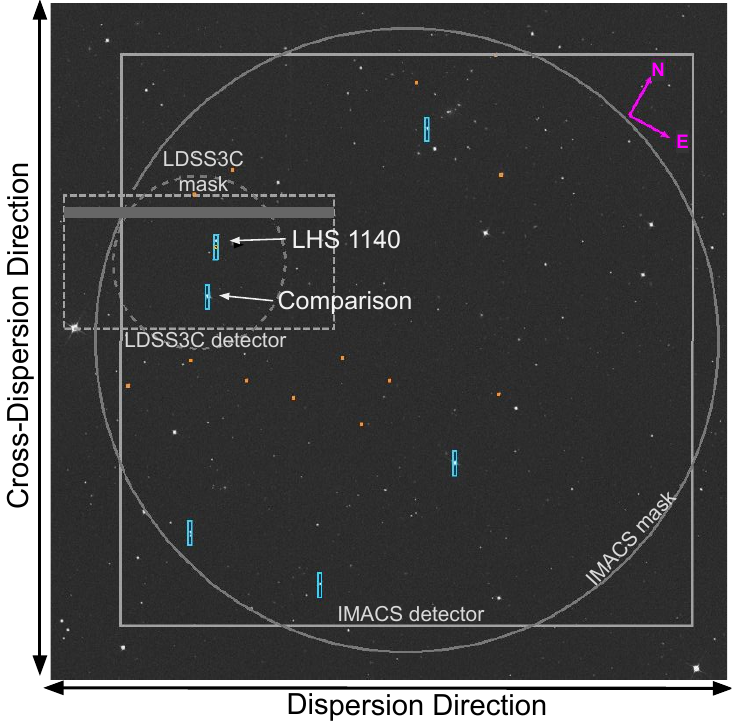}
\caption{On-sky projection of the LHS 1140 field from the Digitized Sky Survey, which is available in \texttt{SAOImageDS9} \citep{Joye2003}. The solid gray circle and square outlines are the mask and detector footprint, respectively, of the IMACS instrument. The dashed gray circle and rectangle outlines are the mask and detector footprint, respectively, of the LDSS3C instrument. Light-blue rectangles are the IMACS science slits; the LHS 1140 and comparison star \mbox{2MASS J00450309--1518437} slits are marked. LHS 1140 is a high-proper-motion star. The yellow circle in the LHS 1140 slit shows the position of LHS 1140 at the time of the 2018 observations (Table~\ref{tab:obs}). We observe four other comparison stars with IMACS but do not use them in the analysis in order to minimize the difference between the IMACS and LDSS3C observations. Orange squares indicate the IMACS alignment holes. There is at least 50$''$ separation in the cross-dispersion direction between the IMACS alignment holes and the science slits in case we needed to model out-of-slit flux (see Section~\ref{sec:analysis}). For LDSS3C, the sizes of the LHS 1140 and comparison star slits are slightly shorter in the cross-dispersion direction than shown (see Section~\ref{subsec:LDSS3Cobs}). For clarity, we do not show the alignment star holes for LDSS3C. The gray-filled strip on the LDSS3C detector indicates a region of bad pixels where slits should not be placed.}
\label{fig:ds9regions}
\end{figure}

When designing these observations, we wanted to keep as many aspects in common as possible between the LDSS3C and IMACS instruments so as to minimize the systematic differences between the two. The field of view of LDSS3C is 8.3$'$, while the f/2 camera on IMACS has a field of view of 30$'$. The field of LHS 1140 is relatively sparse. Fortunately, there is a comparison star, 2MASS J00450309--1518437, located 145.34$''$ away (Figure~\ref{fig:ds9regions}, Table~\ref{tab:stars}). This main-sequence G-type star is nonvariable in the MEarth photometry down to the 1 mmag level (J. Irwin, private communication), and is brighter than LHS 1140. To compare, $T = 11.2991$ for LHS 1140, while $T = 10.5629$ for the comparison star, where $T$ stands for TESS magnitude \citep{Stassun2019}. Because the comparison star is brighter, we are limited by the photon noise of LHS 1140, not the comparison.

To get the same wavelength coverage for LHS 1140 and the comparison star, we ideally want to orient our mask such that the two stars are aligned in the cross-dispersion (spatial) direction. However, there is a background star that was 16.5$''$ away from LHS 1140 during the observations. Lining up LHS 1140 with the comparison star would have placed this background star within a few arcseconds of the edge of the slit. To ensure that this background star did not contaminate the LHS 1140 spectrum by peaking in and out of the slit during observations, we oriented the LDSS3C and IMACS masks such that the spectra of LHS 1140 and the background star are dispersed parallel to each other, with the comparison star almost aligned in the cross-dispersion direction (Figure~\ref{fig:ds9regions}). Because LHS 1140 is a high-proper-motion star it will be necessary to check its position with respect to any background stars in future observations.

\begin{table}
\centering
\caption{Stars Used in This Work \label{tab:stars}}
\begin{tabular}{c|cc}
\tablewidth{0pt}
\hline
\hline
& Target & Comparison\\
\hline
Name & LHS 1140 & 2MASS J00450309--1518437\\
RA  &  00:44:59.33 & 00:45:03.09 \\
Dec & -15:16:17.54 & -15:18:43.87\\
$V$ mag    & 14.15  & 11.01\\
$T$ mag* & 11.2219 & 10.5629\\
$J$ mag    & 9.612   & 9.975\\
Spectral type & M4.5 & G3\\
\hline
\end{tabular}
\begin{minipage}[t]{0.95\linewidth}
\hfill\break
{*The TESS bandpass ranges from 600 - 1000 nm, which is the range over which our observations are made with Magellan Baade/IMACS and Magellan Clay/LDSS3C.}
\end{minipage}
\end{table}

\subsection{Magellan I (Baade) IMACS Observations} \label{subsec:IMACSobs}

The Inamori-Magellan Areal Camera and Spectrograph (IMACS) can perform both imaging and spectroscopy. Its detector is made up of eight CCDs that produce an 8192$\times$8192 pixel mosaic, or $27.5'\times27.5'$ (\href{http://www.lco.cl/telescopes-information/magellan/instruments/imacs/user-manual/the-imacs-user-manual}{IMACS User Manual}). We use the f/2 camera, which has a 30$'$ field-of-view diameter. With this field of view we are able to capture five comparison stars, but we only use 2MASS J00450309--1518437 (Table~\ref{tab:stars}) in the analysis in order to be consistent with the LDSS3C observations.

Between the 2017 and 2018 observations we discovered large instrument systematics that led us to redesign our 2018 mask. These systematics and potential solutions are discussed in detail in Section~\ref{subsec:IMACSextract}, but we present this new mask in Figure~\ref{fig:ds9regions}. The key improvements to the 2018 mask are (1) slits that are 70$''$ long in the cross-dispersion direction in order to estimate the sky background outside of the extended point-spread function of the stellar spectra and (2) ensuring that the area on either side (in the cross-dispersion direction) of the slits has no alignment holes in case we need to model and remove out-of-slit flux. The slit widths in the dispersion direction are 10$''$ to avoid light losses. We recommend that future users of IMACS for similar observations adopt these features when designing their masks. We also cut a calibration mask that is identical to the science mask except with slit widths in the dispersion direction of 0.5$''$. 

For our detector settings we use \texttt{2$\times$2} binning and a \texttt{Fast} readout speed. These settings allow for a readout time of 31.4 s, making the duty cycle for these observations 32.3\%. Gains and readout noise levels for each of the eight IMACS chips can be found in the \href{http://www.lco.cl/telescopes-information/magellan/instruments/imacs/user-manual/the-imacs-user-manual#Mosaic_CCD_Cameras}{IMACS} user manual. During the afternoon prior to observations, we use the science mask to take biases, darks, and quartz flats, and we use the 0.5$''$-slit calibration mask to take helium, neon, and argon arcs. During nighttime observations, we take a nondispersed reference image of the LHS 1140 field with the science mask before and after the science observations. After the nighttime observations we take another set of biases and darks. The 16-bit analog-to-digital converter (ADC) has a saturation limit of 65,535 analog-to-digital units (ADUs), which we do not surpass for all pixels used in the data analysis. We note that with IMACS, the overscan region is sufficient for bias-level subtraction and dark current adds only a few e$^-$ per hour. While biases and darks do not greatly affect our data reduction, taking enough flats is crucial. We were careful to collect at least as many photons in our quartz flats as we do in-transit photons of LHS 1140 in order to not be noise limited by the flats.

For all observations requiring a disperser (i.e., flats, arcs, and science spectra), we use the Gri-300-26.7 grism (300 lines mm$^{-1}$ with a blaze angle of 26.7$^{\circ}$). This grism has a wavelength range of 500-900 nm and a central wavelength of 800 nm. This gives a dispersion of 0.125 nm pixel$^{-1}$. With this grism we use the WBP 5694-9819 order-blocking filter to mitigate any blue light that could cause second-order contamination in our spectra.

\subsection{Magellan II (Clay) LSDD3C Observations} \label{subsec:LDSS3Cobs}

The Low Dispersion Survey Spectrograph (LDSS3C) has gone through several upgrades to make it more sensitive at redder wavelengths. The instrument has a single CCD detector made up of 2048$\times$4096 pixels or $6.4'\times13'$ (\href{http://www.lco.cl/Members/gblanc/ldss-3/ldss-3-user-manual-tmp}{LDSS3C User Manual}). The 8.3$'$-diameter field-of-view radius of LDSS3C means that 2MASS J00450309--1518437 (Table~\ref{tab:stars}) is the only comparison star we are able to observe simultaneously with LHS 1140. We cut our slits 10$''$ wide in the dispersion direction to avoid light losses as seeing and air mass change throughout the night. We cut the comparison star slit 20$''$ long in the cross-dispersion direction in order to capture enough photons to remove the sky background. We cut the LHS 1140 slit 30$''$ longer on one side to account for the background star near LHS 1140. We also cut a mask for wavelength calibrations, which is identical to the science mask except with slit widths of 0.5$''$ in the dispersion direction.

We present the alignment of our science mask on the sky in Figure~\ref{fig:ds9regions}. The LDSS3C detector suffers from some hot pixels, which can saturate and ruin a spectrum. We mark these pixels with a gray-filled rectangle over the LDSS3C detector.

Our detector settings are as follows: \texttt{2$\times$2} detector binning, \texttt{Fast} readout speed, and \texttt{Low} gain. We find that this allows for a 15.6 s readout time, bringing the duty cycle to 46.9\%. Gains and readout noise can be found in the \href{http://www.lco.cl/Members/gblanc/ldss-3/ldss-3-user-manual-tmp#section-12}{LDSS3C} user manual. Note that the \texttt{Low} gain setting actually refers to the inverse gain and therefore allows for longer exposure times than the \texttt{High} gain setting. The full well depth of the detector is 200,000 e$^-$, with a linear pixel response up to 177,000 e$^-$ \citep{Stevenson2016a}. Like IMACS, the 16-bit analog-to-digital converter (ADC) of LDSS3C has a saturation limit of 65,535 analog-to-digital units (ADUs), which we do not surpass for all pixels used in the data analysis.

Using the science mask, we take biases, darks, and quartz flats during the afternoon prior to observations. We also take helium, neon, and argon arcs using the 0.5$''$ calibration mask. During nighttime observations, we take a nondispersed reference image of the LHS 1140 field with the science mask before and after the science observations. After the nighttime observations, we take another set of biases and darks.

For all observations that require a disperser (i.e., flats, arcs, and science spectra) we use the VPH-Red grism, which provides a wavelength coverage of 640-1040 nm (see \citet{Stevenson2016a} for details). The VPH-Red grism has a high throughput at redder wavelengths where LHS 1140, an M star, is brightest. We use the OG590 order-blocking filter to mitigate order contamination introduced to the spectra by the VPH-Red grism. 

\section{Data Extraction} \label{sec:extract}

In this section we discuss how we turn the raw IMACS and LDSS3C data---a time series of \texttt{FITS} files containing 2D stellar spectra---into a time series of 1D stellar spectra, for both LHS 1140 and the comparison star. This final product of the extraction will be the starting point of the data analysis (Section~\ref{sec:analysis}), where we investigate the planet radius of LHS 1140b at different wavelengths. 

The process for extracting the IMACS and LDSS3C spectra of LHS 1140 and the comparison star is identical. We use the custom pipeline \href{http://github.com/zkbt/mosasaurus}{\texttt{mosasaurus}} to perform the extraction. This pipeline has evolved from earlier versions \citep[e.g.,][]{Diamond-Lowe2018} and is now generalized for IMACS and LDSS3C. Though still specialized, this code is modular and may be useful to others performing multiobject transmission spectroscopy of exoplanets.

\subsection{\texttt{mosasaurus} extraction steps}\label{subsec:mosasaurus}
\begin{figure*}[!ht]
\includegraphics[width=1\textwidth]{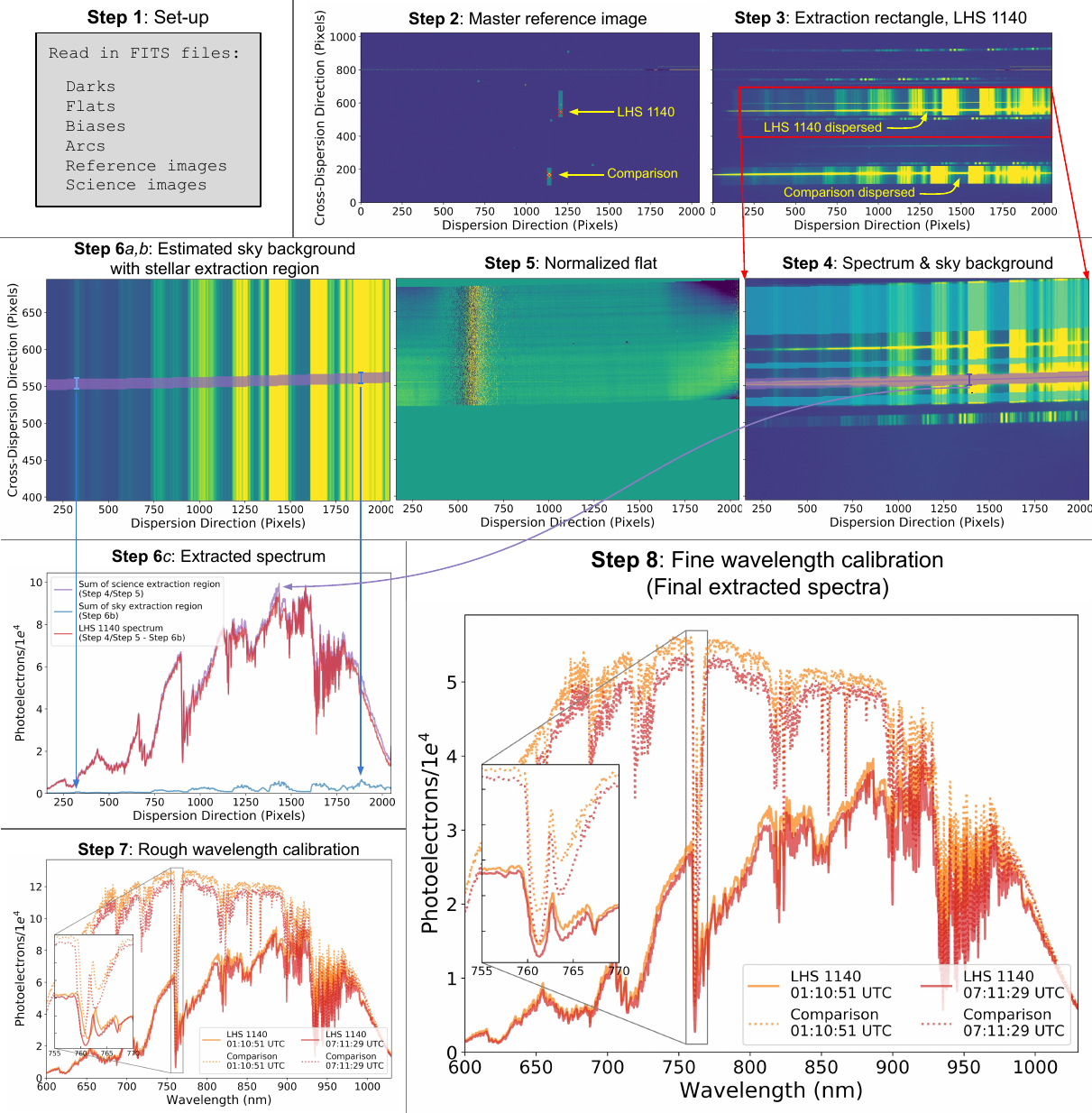}
\caption{Steps of the extraction process performed with the custom \texttt{mosasaurus} pipeline. For a full description of each step, see Section~\ref{subsec:mosasaurus}. These data products are from the 2018 LDSS3C data set.}
\label{fig:extract}
\end{figure*}

Turning raw images into a time series of wavelength-calibrated 1D spectra is a long process. Here we outline the steps of our pipeline. A visual representation of the steps can be see in Figure~\ref{fig:extract}. 

\begin{enumerate}
    \item \textit{Setup.} We read in the \texttt{FITS} files we need for the extraction. These are the darks, biases, quartz flats, arcs (helium, neon, and argon), undispersed reference images, and science images. Following the prescription of \citet{Eastman2010}, we convert the UTC time stamps recorded in the headers of these images into a single BJD$_{TDB}$ time stamp marking the middle of the exposure.
    \item \textit{Master images.} For each type of image we stitch the raw \texttt{FITS} files together to create a coherent image for each of the input files. For IMACS, this results in a 4096$\times$4096 pixel image, and for LDSS3C, a 1024$\times$2048 pixel image (recall that we used 2$\times$2 binning on each instrument). In the process of stitching, we trim the bias overscan regions from each CCD chip (eight for IMACS, two for LDSS3C) and subtract their median in the cross-dispersion direction from the rest of the image. We then take an average of each image type to create the master images. We do this by comparing all of the images of a type and rejecting outliers that deviate by 5$\times$ the median absolute deviation (MAD), and then taking the mean of the images. We refer to this rejection of outliers and averaging of the images as ``stacking.'' Depending on the image type, we perform extra calibrations:
    \begin{enumerate} 
        \item \textit{Biases} We simply stitch and stack all bias images to make the master bias image.
        \item \textit{Darks} We stitch each dark image and then subtract the master bias. Then, we stack the dark images to create the master dark image.
        \item \textit{Flats, arcs, reference images, science images} In the process of stitching these files together, we multiply each CCD chip by the appropriate gain listed in the IMACS and LDSS3C user manuals. After stitching, we subtract the master bias and master dark from each image and then stack each image type to create the master flat, arc, reference, and science images. In Figure~\ref{fig:extract} we show a master reference image, with red crosses marking LHS 1140 and the comparison star in their slits. From the master flat we also create a bad pixel mask.
    \end{enumerate}
    \item \textit{Extraction rectangles.} Using an interactive plotting tool developed for \texttt{mosasaurus}, we indicate which stars on the master reference image we wish to extract. \texttt{mosasaurus} then cuts out a rectangle around each of the desired spectra on the master science image and a corresponding rectangle from the master flat and arc. The extraction rectangle for the LHS 1140 spectrum is shown in red in Figure~\ref{fig:extract}.
    \item \textit{Stellar spectra and sky-background.} Using the rectangle cut from the master science image, we use an interactive plotting tool to indicate the spectral traces of LHS 1140 (purple line, Figure~\ref{fig:extract}) and the comparison star. An extraction region is defined as a set number of pixels away from the center of the stellar trace (purple band). We also indicate portions of sky background on either side of the spectral trace (light-blue bands). These are used to fit and remove the sky-background flux from the stellar flux during extraction. 
    \item \textit{Normalized flat for each star.} We use the extraction rectangles cut from the master flat to create a normalized flat for each star (flat for LHS 1140 shown in Figure~\ref{fig:extract}). The normalized flat is made by dividing each column of pixels in the cross-dispersion direction by the median value of that column. When making the median filter, we only use portions of the flat extraction rectangle that correspond to pixels that are included in the spectral extraction, i.e., the stellar extraction region, the sky-background regions, and any intervening regions. We divide the extraction rectangles for each science exposure by the corresponding normalized flat.
    \item \textit{Extract spectra.} We cycle through the science exposures and extract spectra of LHS 1140 and the comparison star in the following steps:
    \begin{enumerate}
        \item \textit{Sky background.} For each column of pixels in the cross-dispersion direction of an extraction rectangle we use the sky-background regions (designated in Step 4) to make a second-order polynomial fit to the pixel column. This makes a 2D, polynomial-smoothed estimate of the sky background in the extraction rectangles of each exposure (Figure~\ref{fig:extract}). We note that a median of the sky-background pixels can also be used, with similar results.
        \item \textit{Sky in stellar extraction region.} We take the portion of the 2D sky background that covers the stellar extraction region designated in Step 4 (purple) and sum in the cross-dispersion direction, creating a 1D estimate of the sky background (light-blue spectrum in Figure~\ref{fig:extract}).
        \item \textit{Extracted spectrum.} We divide the extraction rectangle (Step 4) by the normalized flat (Step 5) and sum the stellar extraction region in the cross-dispersion direction (purple spectrum in Figure~\ref{fig:extract}). We then subtract the 1D sky-background estimate (blue spectrum) to get the extracted spectrum (red spectrum).
    \end{enumerate}
    \item \textit{Rough wavelength calibration.} We need to create a wavelength solution to convert the extracted spectra from flux versus pixel to flux versus wavelength. Using another interactive plotting tool, we take the arc extraction rectangles for each star and mark the helium, neon, and argon lines. We then compare where our marked wavelengths are in pixel space to a template of lines for the grisms we used with the \href{http://www.lco.cl/telescopes-information/magellan/instruments/ldss-3/atlas-of-comparsion-lamp-spectra-for-ldss-3/lamp-spectra}{LDSS3C} and \href{https://github.com/zkbt/mosasaurus/blob/master/data/IMACS/gri-300-26.7/HeNeAr.txt}{IMACS} detectors. We fit the marked arc lines to the template lines using Legendre polynomials and apply this wavelength solution to each of the extracted spectra. Finally, we resample each spectrum so that they are on a common, uniform wavelength grid; we ensure that flux is conserved in this process. We enforce flux conservation by performing the interpolation on the cumulative distribution function of the flux, thereby ensuring that an integration of the flux over any wavelength range returns the same value as it would have had we not resampled the spectrum. The result works reasonably well, but there are visible mismatches in spectral features between LHS 1140 and the comparison star, and also between exposures taken at different times throughout the night (zoomed-in inset, Figure~\ref{fig:extract}). This rough wavelength calibration aligns the spectra to within 0.5 nm for IMACS spectra and 1.0 nm for LDSS3C spectra (0.2 and 0.4 pixels, respectively). We will eventually bin these spectra into 20 nm wavelength bins, and this slight misalignment can introduce additional noise.
    \item \textit{Fine wavelength calibration.} For a single spectrum we isolate prominent telluric and stellar spectral features---the O$_2$ doublet (760.5 nm), the Ca triplet (849.8, 854.2, and 866.2 nm), and the water line forest (930-980 nm)---and cross-correlate them with the same features in all other spectra in a data set. Our stars are close enough (in the Sun's local moving group) and our spectral resolution low enough (upper limits of 250 km s$^{-1}$ pixel$^{-1}$ for IMACS and 165 km s$^{-1}$ pixel$^{-1}$ for LDSS3C) that comparing telluric O$_2$ and H$_2$O features to stellar Ca features is not introducing errors into our wavelength calibration. After the cross-correlation, we rerun the flux-conserving resampling routine to reflect the new wavelength grid for each spectrum. With this technique we align our spectra to within 0.25 nm (or 0.10 pixels; zoomed-in inset, Figure~\ref{fig:extract}). We use multiple data sets for this work, so we also wavelength-calibrate between the data sets.
\end{enumerate}

We note that one improvement to our pipeline would be to change the extraction region around the stellar spectra such that it evolves over the time series. This would entail retracing the stellar spectra in every exposure \citep{Jordan2013,Rackham2017,May2018} or utilizing an optimal extraction routine \citep{Stevenson2016a, Bixel2019}. Systematics introduced by using a fixed aperture are decorrelated against during analysis (Section~\ref{sec:analysis}) and do not alter the results of this work.

\begin{figure*}
\includegraphics[width=1\textwidth]{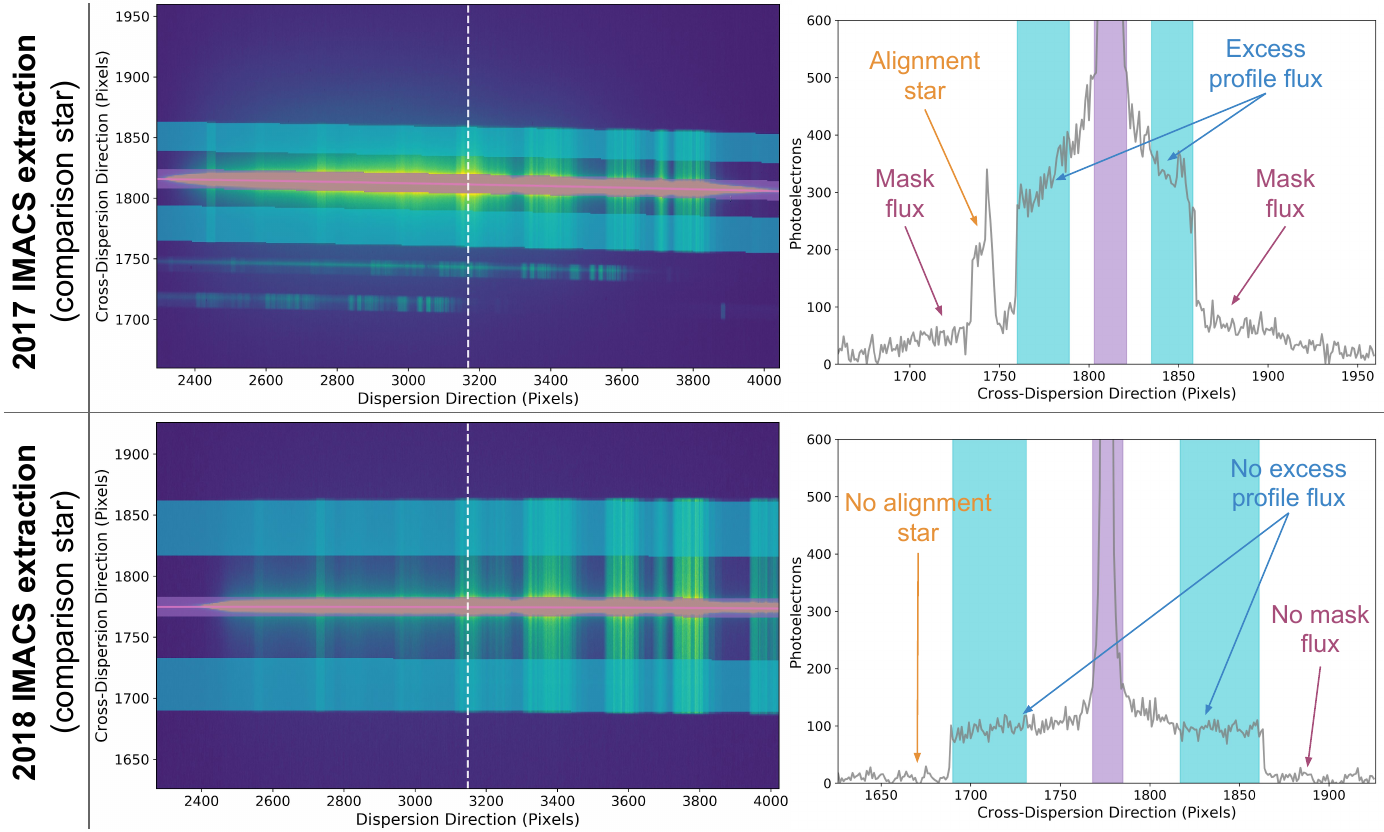}
\caption{Right: comparison star extraction rectangles for exposures from the 2017 and 2018 IMACS and data sets (similar to Step 4 in Section~\ref{subsec:mosasaurus}), as well as a cut in the cross-dispersion direction (white dashed line). Exposure times are 15 s for both observations. Left: cut of flux profiles in the cross-dispersion direction. The 2017 IMACS profile exhibits excess ``mask flux,'' as well as excess flux in the wings of the stellar profile. Light-blue and purple bands correspond to the bands in the extraction rectangles; they indicate which points in the profile are used to estimate the sky background (light blue) and which points are summed to extract the stellar spectrum (light purple).}
\label{fig:IMACSextract}
\end{figure*}

\subsection{Issues with Magellan I (Baade) IMACS data} \label{subsec:IMACSextract}

The 2017 IMACS data set exhibited anomalies that led us to perform a deep exploration of this data set and ultimately decide not to include it in our analysis. The ACCESS Collaboration \citep{ACCESSCollab2014} noticed similar systematics, which are thoroughly outlined in \citet{Espinoza2017}\footnote{\href{https://repositorio.uc.cl/handle/11534/21313}{repositorio.uc.cl/handle/11534/21313}}.

We find that the source of these anomalies is an excess of light scattered by the IMACS instrument that occurs when the disperser is in place \citep[][Chapter 3]{Espinoza2017}. Figure~\ref{fig:IMACSextract} shows that this excess light adds nonnegligible flux in portions of the detector that should be masked. We call this excess flux, unimaginatively, ``mask flux.'' We also see an excess of flux in the wings of the stellar profile in the cross-dispersion direction. In Figure~\ref{fig:IMACSextract} we show the extraction rectangle of the comparison star from the 2017 IMACS data set, as well as a cut across the extraction rectangle in the cross-dispersion direction, to demonstrate the excess flux that we see. We compare these to the same figures for the 2018 IMACS data set, which does not exhibit excess flux. 

\citet{Espinoza2017} outlines a process to model and remove the mask flux. We were able to remove the mask flux from the comparison star spectra; however due to the alignment star holes near the LHS 1140 slits and the closeness of LHS 1140 to the edge of the slit, the flux profile in the cross-dispersion direction is difficult to model for this star. We therefore do not include the 2017 IMACS data set in our analysis.

For the 2018 IMACS data set we made significant changes to our mask (see Section~\ref{subsec:IMACSobs}) to ensure that we captured the full PSF of LHS 1140 and the comparison star and were able to model and remove the mask flux. The 2018 observations occurred on a dark night (no moon), and we did not see the same excess mask flux in these data. The extralong slits in the cross-dispersion direction did help us to capture the full PSF of LHS 1140 and the comparison star, along with enough sky background to do the extraction.

\section{Data Analysis} \label{sec:analysis}

In Section~\ref{sec:extract} we turned the raw \texttt{FITS} files that we collected during our observations into time series of 1D wavelength-calibrated spectra of LHS 1140 and the comparison star. These time series spectra exhibit two types of systematic trends that we address before constructing a transmission spectrum: (1) instrument systematics from the Magellan telescopes and the IMACS and LDSS3C spectrographs, and (2) telluric systematics from Earth's atmosphere, which we peer through as we observe. So as not to tamper with the transit information buried in the time series, we model the systematics at the same time as we model the transit properties of LHS 1140b. We ultimately want to simultaneously analyze the spectra from each data set in order to construct the transmission spectrum.

We built a custom data analysis pipeline that picks up where \texttt{mosasaurus} left off. The pipeline, named \href{https://github.com/hdiamondlowe/decorrasaurus/releases/tag/v2.0}{\texttt{decorrasaurus,\footnote{This pipeline is the new and improved cousin of the \href{https://github.com/hdiamondlowe/detrendersaurus/releases/tag/v1.0}{\texttt{detrendersaurus}} pipeline, which is no longer used.}}} is built to take in IMACS and LDSS3C data cubes from \texttt{mosasaurus} and return systematic-removed light curves that can be turned into transmission spectra. The \texttt{decorrasaurus} pipeline supports two methods of treating these effects in the light curves: (1) with a linear fit, and (2) with a Gaussian process (GP) regression. Where the methods differ in the analysis, we split the steps into a part A and a part B, respectively.

\subsection{\texttt{decorrasaurus} decorrelation steps}\label{subsec:decorrasaurus}

\begin{table*}[ht]
\centering
\caption{Systematic Model Vectors\label{tab:sysparams}}
\begin{tabular}{lp{10.5cm}ccccccc}
\tablewidth{0pt}
\hline
\hline
Input &  & Is a & \multicolumn{6}{c}{Use in Data Sets\tablenotemark{b}}\\
\cline{4-9}
Vector & \centering{Input Vector Definition} & Function &  \multicolumn{2}{c}{L17} & \multicolumn{2}{c}{I18} & \multicolumn{2}{c}{L18}\\
\rowcolor{white} Name && of\tablenotemark{a}& A & B & A & B & A & B\\
\hline
Time  & From the header files, the average time from the start of the exposure to the end of the CCD readout for that exposure. & t & ... & \checkmark & ... & \checkmark & ... & \checkmark\\
Air mass  & From the header files, the average air mass of the field recorded at each exposure during observation. & t &  \checkmark &  \checkmark & \checkmark & \checkmark & \checkmark & \checkmark \\
Rotation angle & From the header files, rotation angle of the instrument recorded at each exposure. This can be correlated with changes in illumination or flexure during observation. & t &  \checkmark & \checkmark  & \checkmark & \checkmark & \checkmark & \checkmark\\
Centroid & Derived during extraction, the stellar centroid measured in the cross-dispersion direction. This is the median of the centroids across all wavelengths for each star in each exposure. & t, s & & & & & \\
Width & Derived during extraction, the width of the spectral trace in the cross-dispersion direction. This is the median of the measured widths across all wavelengths for each star in each exposure. & t, s & \checkmark &  & \checkmark &  & \checkmark &\\
Peak  & Derived during extraction, the brightness of the brightest pixel in the cross-dispersion direction measured at every wavelength for each star in each exposure. This is summed in wavelength space for each wavelength bin. & t, s, $\lambda$ & & & \checkmark & & \checkmark & \\
Shift  & Derived during extraction, the linear change in the dispersion direction needed to align the spectra with each other. This is calculated for each star in each exposure. & t, s & \checkmark &  & \checkmark &  & \checkmark &  \\
Stretch  & Derived during extraction, the multiplicative change in the dispersion direction needed to align the spectra with each other. This is calculated for each star in each exposure. & t, s & & & && \\
Polynomial & Specified during analysis, the degree of the Legendre polynomial component of the model. & t & 3 & ... & 2 & ... & 3 & ...\\
\hline
\end{tabular}
\begin{minipage}[t]{0.95\linewidth}
\hfill\break
{\tablenotetext{a}{Vectors can be functions of time $t$, star $s$, and wavelength $\lambda$}
\tablenotetext{b}{Data sets: L17 = LDSS3C 2017, I18 = IMACS 2018, L18 = LDSS3C 2018; Modeling method: A = linear, B = Gaussian process\\ The time parameter is only available for the GP method, while the Legendre polynomial set is only available in the linear method}}
\end{minipage}
\end{table*}

Turning time series of wavelength-calibrated 1D spectra into decorrelated light curves and a transmission spectrum is also a lengthy process. Here we outline the steps of our pipeline.

\begin{enumerate}
    \item \textit{Setup.} We read in the \texttt{mosasaurus} data cubes that we wish to analyze. \texttt{decorrasaurus} can work with a single data set, or multiple data sets simultaneously if parameters are to be jointly fit across multiple data sets. We also specify which parameters should be fixed or varied and how to bin the light curves in wavelength space. At this stage we specify which transit parameters and which systematic parameters we will fit. The input vectors, associated with the systematic parameters, are normalized by subtracting the mean and dividing by the standard deviation for each vector.
    \item \textit{Make light curves.} Here we transform a time series of wavelength-calibrated 1D spectra of LHS 1140 and the comparison star into a time series of normalized fluxes, or a light curve. This requires summing up each of the spectra in a given wavelength bin. We chop the spectra in wavelength space (recall that all spectra were interpolated onto a common wavelength grid in Step 8 of Section~\ref{subsec:mosasaurus}) in order to make the wavelength bins. If necessary, we take fractions of pixels in order to meet the chosen wavelength cutoffs. We normalize each wavelength-binned time series of fluxes by the median flux for that time series. We then divide the LHS 1140 time series by the comparison star time series to make the light curve.
    \item [3A.] \textit{Model the data: linear.} The linear model assumes that the noise in the light curves is Gaussian. We construct a model to the data that has two components:
    \begin{enumerate}
        \item \textit{Systematics.} The systematics component of the model $\mathcal{S}(t)$ is composed of a Legendre polynomial specified during setup and the input vectors recorded from the data extraction. Table~\ref{tab:sysparams} lists the input vectors used in the systematics model, along with explanations. In the linear model, we decorrelate the light curves against these input vectors. This model component can be described as
        \begin{equation}
            \mathcal{S}(t) = 1 + \sum_{n=1}^{N_{\mathrm{poly}}}c_nP_{n-1}(t) + \sum_{m=1}^{M_{\mathrm{phys}}}c_mR_m(t,^*\!\!\lambda)
        \end{equation}
        where $t$ is the time array covered by the light curve, $P_{n-1}$ are the set of Legendre polynomials, $R_m(t,^*\!\!\lambda)$ are the input vectors derived from the extraction (they are all functions of time $t$ but some also have a wavelength $\lambda$ dependency), and $c_n,c_m$ are the coefficients we fit for.
        \item \textit{Transit.} The transit component of the model $\mathcal{T}(t)$ is made with the \texttt{batman} package \citep{Kreidberg2015}. Table~\ref{tab:transitparams} explains which transit parameters we fix or vary for each fit we perform.
    \end{enumerate}
    The complete model $\mathcal{M}(t)$ that we fit to the light curve is
    \begin{equation}
        \mathcal{M}(t) = \mathcal{S}(t)\mathcal{T}(t)
    \end{equation}
    In Steps 5 and 6 we fit for the systematics coefficients and the transit parameters simultaneously to achieve the best fit to the light-curve data.
    \item [3B.] \textit{Model the data: GP.} Using the open-source package \texttt{george} \citep{Foreman-Mackey2015}, we implement a GP regression to model the data. Now, we assume that noise in the data can be correlated as well as Gaussian. Again, there are two components that go into the model:
    \begin{enumerate}
        \item \textit{Noise.} Following a long line of work using a GP regression to model light curves \citep[e.g.,][]{Gibson2012,Gibson2014,Evans2017,Kirk2019}, we use a covariance function, or kernel, to model the data. We do this by computing the covariance matrix using the input vectors. Here, we do not directly decorrelate the light curves against the input vectors; rather, we use these vectors to construct a covariance matrix that is used to predict the noise in the light curves. We use a Mat\'ern $3/2$ kernel of the form
        \begin{equation}
        k(r^2) = A^2\left(1 + \sqrt{3\frac{r^2}{L^2}}\right)e^{-\sqrt{3\frac{r^2}{L^2}}}
        \end{equation}
        where $A^2$ is the amplitude, $L$ is the length scale, and $r$ is the distance between two points in the light curve. In practice, the amplitude $A^2$ multiplies a summation of Mat\'ern $3/2$ kernels, each of which corresponds to an input vector. $A^2$ itself is a constant kernel in the \texttt{george} framework. We set the length scale $L$ as the hyperparameter of each kernel, which describes how correlated the light-curve points are, given the input vector. The hyperparameter of a kernel has a physical meaning in that it determines how close points in the input vectors have to be in order to be correlated with each other. If the length scale is too small, the GP will try to fit every single point, thereby overfitting the data. We choose to use Mat\'ern 3/2 kernels with the assumption that in the light curves, points closer together in time are more likely to be correlated than those farther away. In this regard the functional form of a Mat\'ern 3/2 kernel is similar to that of a squared exponential kernel. We also include a white-noise component, which is an optional input in the \texttt{george} implementation of the GP, and is an additive parameter that accounts for the Gaussian and independent nature of the uncertainties associated with counting photons.
        \item \textit{Transit.} As in the linear method, we use the \texttt{batman} package \citep{Kreidberg2015} to construct the transit. Table~\ref{tab:transitparams} explains which transit parameters we fix or vary for each final fit we perform. This transit model is set as the mean, around which the data deviates, in the \texttt{george} Gaussian process object.
    \end{enumerate}
    In Step 5B we fit for the kernel hyperparameters but keep the transit parameters (the mean of the GP) fixed. In Step 6B we fit for the kernel hyperparameters and the transit parameters simultaneously to achieve the best fit to the light-curve data.
    \item [4.] \textit{Limb-darkening coefficients.} The parameters that describe the opacity at the stellar limb are crucial for constructing an accurate transit model. Following the conclusions of \citet{Espinoza2016} regarding the optimal limb-darkening model for stars with effective temperatures $T_{\mathrm{eff}} < 4000$ K, we employ a logarithmic limb-darkening law, which is supported by \texttt{batman}. We use the \texttt{Limb Darkening ToolKit} \citep[\texttt{LDTk;}][]{Parviainen2015} to interpolate stellar models from the \texttt{PHOENIX} library \citep{Husser2013} and calculate the logarithmic limb-darkening coefficients for the wavelength range of interest. (The \texttt{ldtk} package does not include the logarithmic law, but it is simple to add additional laws to the package; it is the interpolation of the stellar models that is important.) In order to decorrelate the limb-darkening coefficients, we reparameterize $l_0$ and $l_1$ returned by \texttt{LDTk} to $q_0 = (1 - l_1)^2$ and $q_1 = (1 - l_0)/(1 - l_1)$ \citep[Equations 2-5,][]{Espinoza2016}. With this reparameterization we can sample the space of physically plausible logarithmic limb-darkening coefficients by sampling $q_0$ and $q_1$ uniformly between (0, 1) when we do a full sampling of the parameter space in Step 6. This is the most conservative choice we could make for the priors when fitting for the limb-darkening parameters and likely degraded our precision on $R_p/R_s$.
    \item [5A.] \textit{Simple minimization: Linear.}  We perform three iterations of a Levenberg--Marquardt least-squares fitting routine using the \texttt{lmfit} package \citep{Newville2016}. The advantage of this fit is that it is fast, which means that we can test different systematic parameters and choose the best ones to marginalize over. 
    \begin{enumerate}
        \item \textit{Iteration 1.} We use the calculated photon noise for each data set to weight the residuals in the least-squares minimization.
        \item \textit{Iteration 2.} We clip any points that are 5$\times$ the median absolute deviation of the residuals. We again use the calculated photon noise to weight the residuals.
        \item \textit{Iteration 3.} We calculate the standard deviation of the residuals for each data set in the fit. We use this calculated error to weight the residuals in the least-squares minimization.
    \end{enumerate}
    After the three iterations, we use the best-fit values to compare different models to each other, employing both the Bayesian information criterion \citep[BIC;][]{Schwarz1978} and the Akaike information criterion \citep[AIC;][]{Akaike1998}, which penalize excessive model parameters. When we have found the best input vectors to decorrelate against (Table~\ref{tab:sysparams}), we continue to the next step, where we more fully sample the parameter space in order to estimate the uncertainties of the free parameters. 
    \item [5B.] \textit{Simple minimzation: GP.} We perform two iterations of a minimizing routine using \texttt{scipy.optimize} and the \texttt{L-BFGS-B} method, which allows for the computation of a gradient vector, as well as bounds on the input variables.
    \begin{enumerate}
        \item \textit{Iteration 1.} We construct the GP object using \texttt{george}. When pre-computing the GP object, it is necessary to provide uncertainties in the data. We use the calculated photon noise to create the GP object. We then compute the conditional predictive distribution of the GP model on the light curve. This does not yield an optimal fit but does allow us to clip any data points 5$\times$ the median absolute deviation of the residuals. We then perform the minimization.
        \item \textit{Iteration 2.} We calculate the standard deviation of the residuals for each data set and then reconstruct the GP object using these calculated uncertainties to pre-compute the GP object. We then perform the minimization.
    \end{enumerate}
    We do not allow the transit parameters to vary at this stage, making the minimization computationally efficient. The hyperparameters we find from this minimization are used as the starting values for the hyperparameters in the full exploration of the parameter space (Step 6). The task of which input vectors to include in computing the covariance matrix is left until Step 6.
    \item [6.] \textit{Dynamic nested sampling.} To estimate the uncertainties in our free parameters, we need to perform a more complete exploration of the parameter space. This has frequently been done with a Markov Chain Monte Carlo algorithm, such as the one provided in the \texttt{emcee} package \citep{Foreman-Mackey2013}. Here we employ \texttt{dynesty} \citep{Speagle2020}, an open-source dynamic nested sampling routine. \texttt{dynesty} samples all parameters between (0, 1) and requires a prior transform function to translate these values into the real parameter space. It is therefore necessary to provide priors for all parameters in the fit.
    \item [6A.] \textit{Dynamic nested sampling: linear.} We use the $1\sigma$ uncertainties derived from the Levenberg--Marquardt fits (Step 5A) to set the priors for \texttt{dynesty}. For all free parameters we use flat priors bounded at 10$\times$ the Levenberg--Marquardt uncertainties, except for the limb-darkening parameters, where we use a Gaussian prior with the standard deviation equal to 1$\times$ the uncertainty calculated by \texttt{LDTk} (Step 4). We assume that our data are drawn from an uncorrelated Gaussian distribution, but we multiply the standard deviation $\sigma$ by a scaling parameter $s$ to account for excess noise that our model does not capture \citep[see][for a detailed explanation]{Berta2012a}. Our modified log-likelihood function is
    \begin{equation}
        \mathrm{ln}\mathcal{L} = N\mathrm{ln}s - \frac{1}{2s^2}\chi^2 + \mathrm{constant}
    \end{equation}
    where
    \begin{equation}
        \chi^2 = \sum_{i=0}^{N} \left(\frac{d_i - m_i}{\sigma_i}\right)^2
    \end{equation}
    where $N$ is the number of data points, $d$ is the data, and $m$ is the model.
    \item [6B.] \textit{Dynamic nested sampling: GP.} With \texttt{dynesty} we can sample both the posterior distribution and the Bayesian evidence, which we use for model comparison to determine the best input vectors to use to compute the covariance matrix (Table~\ref{tab:sysparams}). For the transit parameters we set uniform priors. We fit the amplitude, kernel hyperparameters, and the white noise with log-uniform priors. We set the priors as follows
    \begin{enumerate}
        \item \textit{Amplitude.} The amplitude is described by a constant kernel which multiplies all of the kernels associated with the noise parameters. We set the lower and upper bounds as
        $$\mathcal{U}(ln(0.01\sigma^2),\ ln(100\sigma^2))$$
        where $\sigma^2$ is the variance of the light curve.
        \item \textit{Kernel hyperparameters.} We fit for the hyperparameter in each of the kernels associated with the input vectors. We set the lower and upper bounds as
        $$\mathcal{U}(ln(\delta d),\ ln(\Delta d))$$
        where $\delta d$ is the absolute value of the average difference between consecutive points in the input vector and $\Delta d$ is 3$\times$ the total range of the input vector (i.e., the maximum value minus the minimum value). Recall that the input vectors were normalized in Step 1.
        \item \textit{White noise.} The white-noise parameter accounts for the Poisson statistics associated with measuring stellar flux. We set the lower and upper bounds of this additive parameter as
        $$\mathcal{U}(ln((250\ \mathrm{ppm})^2),\ ln((2500\ \mathrm{ppm})^2))$$
        where some guess-and-check work went into choosing this range such that it was broad enough to not bias the fit, but narrow enough to explore the parameter space efficiently.
    \end{enumerate}
\end{enumerate}

\subsection{Two data analyses}\label{subsec:3analyses}

\begin{table*}[ht]
\centering
\caption{Transit Model Parameters Used in Final Analyses\label{tab:transitparams}}
\begin{tabular}{lcccc}
\tablewidth{0pt}
\hline
\hline
\multirow{2}{*}{Parameter Name} & \multicolumn{2}{c}{Data Sets Fit Independently} & \multicolumn{2}{c}{Data Sets Fit Jointly}\\
\cline{2-3}\cline{4-5}
& White Light Curve &  $\lambda$-binned Light Curves & White Light Curve & $\lambda$-binned Light Curves\\
\hline
Midtransit time difference, $\Delta t_0$ & Free & Fixed & Free & Fixed\\
Planet-to-star radius ratio, $R_p/R_s$ & Free & Free & Free, joined & Free, joined\\
Limb darkening, [$q0, q1$] & Free & Free & Free, joined & Free, joined\\
\hline
\end{tabular}
\begin{minipage}[t]{0.95\linewidth}
\hfill\break
{\textbf{Note.} We assume a circular orbit for all fits.\\
Free = parameter is allowed to vary.\\
Fixed = parameter is fixed.\\
Free, joined = parameter is allowed to vary, but must be the same for all data sets.\\
Fixed, joined = parameter is fixed and the same for all data sets.}
\end{minipage}
\end{table*}

\begin{figure}
\includegraphics[width=0.5\textwidth]{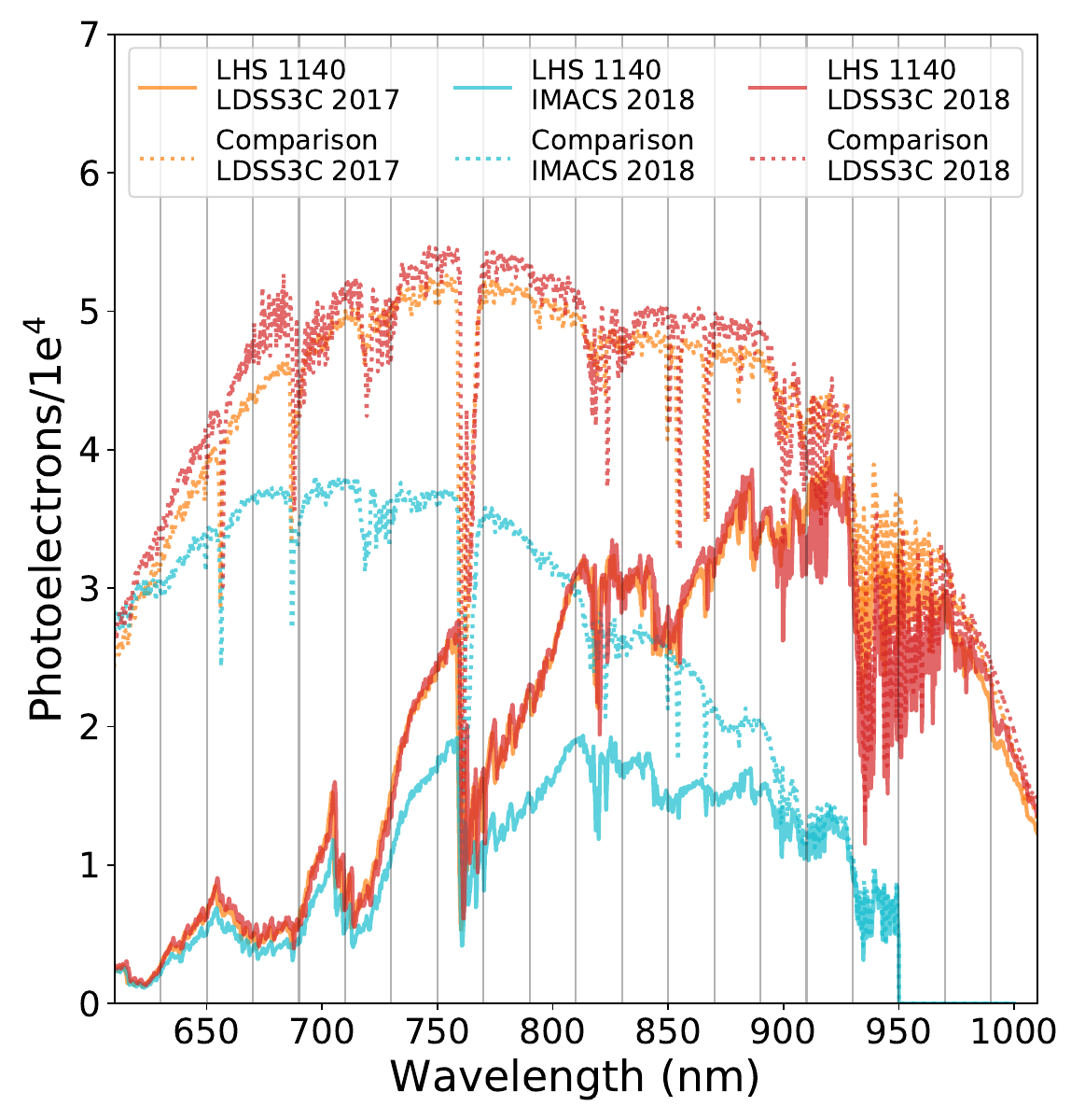}
\caption{Representative spectra of LHS 1140 (solid lines) and the comparison star (dotted lines) from the three data sets we analyze in this work. Gray vertical lines indicate the 20 nm wavelength bins that we use to construct the transmission spectrum.}
\label{fig:spectrabinned}
\end{figure}

\begin{table*}[ht]
\centering
\caption{Transit and Systematic Model Parameters for the Jointly Fit White Light Curve, Gaussian Process Regression \label{tab:whitelcvalues}}
\begin{tabular}{lcccc}
\hline
\hline
\multirow{2}{*}{Parameter Name} & Initial Value & Fitted Value & Priors for & Fitted Value \\ 
        & for Step 5 & from Step 5 &  Steps 5 and 6 & from Step 6  \\
\hline
$\Delta t_0$\tablenotemark{\textsuperscript{1}}, L17 (days)& 0.0 & ... & $\mathcal{U}$(-0.005, 0.005) & -0.0027 $\pm$ 0.00024 \\
$\Delta t_0$\tablenotemark{\textsuperscript{1}}, I18 (days)& 0.0 & ... & $\mathcal{U}$(-0.005, 0.005) & 0.0022 $\pm$ 0.00020 \\
$\Delta t_0$\tablenotemark{\textsuperscript{1}}, L18 (days)& 0.0 & ... & $\mathcal{U}$(-0.005, 0.005) & 0.0023 $\pm$ 0.00022 \\
$R_p/R_s$ & 0.074 & ... & $\mathcal{U}$(0.055, 0.085) & 0.073 $\pm$ 0.0023 \\

$q_0$ & 0.36  & ... & $\mathcal{U}$(0, 1) & 0.47 $\pm$ 0.26 \\
$q_1$ & 0.39 & ... & $\mathcal{U}$(0, 1) & 0.65 $\pm$ 0.13 \\ \addlinespace[8pt]

ln(amplitude), L17 & -11.7 & -15.7 & $\mathcal{U}$(-17.0, -7.0) & $-15.6^{+0.57}_{-0.42}$ \\
ln(time), L17 & 2.05 & 2.05 & $\mathcal{U}$(-5.39, 2.34) & $-3.95^{+0.78}_{-0.60}$ \\
ln(airmass), L17 & 2.25 & 2.53 & $\mathcal{U}$(-6.91, 2.54) & $2.01^{+0.39}_{-0.87}$  \\
ln(rotation angle), L17 & 1.76 & -6.10 & $\mathcal{U}$(-6.10, 2.05) & $1.40^{+0.48}_{-1.1}$ \\
ln(white noise), L17 & -15.2 & -16.9 & $\mathcal{U}$(-16.6, -12.0) & $-16.5^{+0.10}_{-0.053}$ \\ \addlinespace[8pt]

ln(amplitude), I18 & 11.8 & -14.8 & $\mathcal{U}$(-17.1, -7.15) & $-12.6^{+0.84}_{-0.82}$ \\
ln(time), I18 & 2.05 & -3.53 & $\mathcal{U}$(-5.00, 2.34) & $0.37^{+1.4}_{-0.95}$ \\
ln(airmass), I18 & 2.22 & 2.51 & $\mathcal{U}$(-5.70, 2.51) & $2.17^{+0.26}_{-0.54}$ \\
ln(rotation angle), I18 & 1.80 & 2.09 & $\mathcal{U}$(-5.61, 2.09) & $1.55^{+0.36}_{-0.65}$ \\
ln(white noise), I18 & -15.2 & -15.1 & $\mathcal{U}$(-16.2, -11.3) & $-16.1^{+0.059}_{-0.029}$  \\ \addlinespace[8pt]

ln(amplitude), L18 & -11.8 & -14.8 & $\mathcal{U}$(-17.1, -7.2) & $-11.3^{+0.87}_{-1.2}$ \\
ln(time), L18 & 2.05 & -3.87 & $\mathcal{U}$(-5.28, 2.34) & $1.4^{+0.72}_{-1.3}$ \\
ln(airmass), L18 & 2.19 & 0.80 & $\mathcal{U}$(-5.70, 2.47) & $1.62^{+0.62}_{-0.92}$ \\
ln(rotation angle), L18 & 1.88 & 2.17 & $\mathcal{U}$(-5.74, 2.17) & $1.58^{+0.46}_{-0.96}$ \\
ln(white noise), L18 & -15.2 & -16.9 & $\mathcal{U}$(-16.6, -11.0) & $-16.5^{+0.11}_{-0.61}$ \\
\hline
\end{tabular}
\begin{minipage}[t]{0.72\linewidth}
\hfill\break
{L17 = LDSS3C 2017; I18 = IMACS 2018; L18 = LDSS3C 2018\\
\textbf{Note.} The steps referenced in the columns refer to those in Section~\ref{subsec:decorrasaurus}. For the purpose of this study, we assume a circular orbit for LHS 1140b.
\tablenotetext{\textsuperscript{1}}{The measured transit midpoint time for each data set can be calculated as $t_0 = T_0 + nP + \Delta t_0$, where ephemeris $T_0 = 2456915.71154 \pm 0.00004$ days and period $P = 24.736959 \pm 0.000080$ \citep{Ment2019}. For the LDSS3C 2017 data set $n = 46$, and for the IMACS and LDSS3C 2018 data sets $n = 61$.}}
\end{minipage}
\end{table*}
In order to marginalize over the appropriate parameters, we build up to the transmission spectrum by first analyzing our data sets separately and then analyzing them jointly. Each analysis involves constructing both a white light curve and a set of wavelength-binned light curves. Table~\ref{tab:transitparams} lists the transit parameters and whether they are free or fixed in each fit. The breakdown of the white light curves into wavelength-binned light curves (20 nm bins) is shown in Figure~\ref{fig:spectrabinned}. For all analyses we assume a circular orbit for LHS 1140b.

\subsubsection{Data sets fit independently}\label{subsubsec:individual}
We first treat our data sets independently. The main purpose of this step is to decide which input vectors should be used in the modeling of each light curve. Table~\ref{tab:sysparams} lists all possible input vectors along with an explanation of how they are constructed, and which are used in the analysis of each light curve in both the linear and GP modeling methods. We use the white light curves to determine the best input vectors to use for each data set. We use the same input vectors in the wavelength bins of a given data set as we do for the white light curve. We ultimately focus on the results from the GP method, and we provide the derived parameters from the simple minimization and full sampling of the parameter space (Steps 5B and 6B of Section~\ref{subsec:decorrasaurus}), along with the priors for the white light curves. We use the same priors for the wavelength bins.

\subsubsection{Data sets fit jointly}\label{subsubsec:joint}
\begin{figure}
\includegraphics[width=0.5\textwidth]{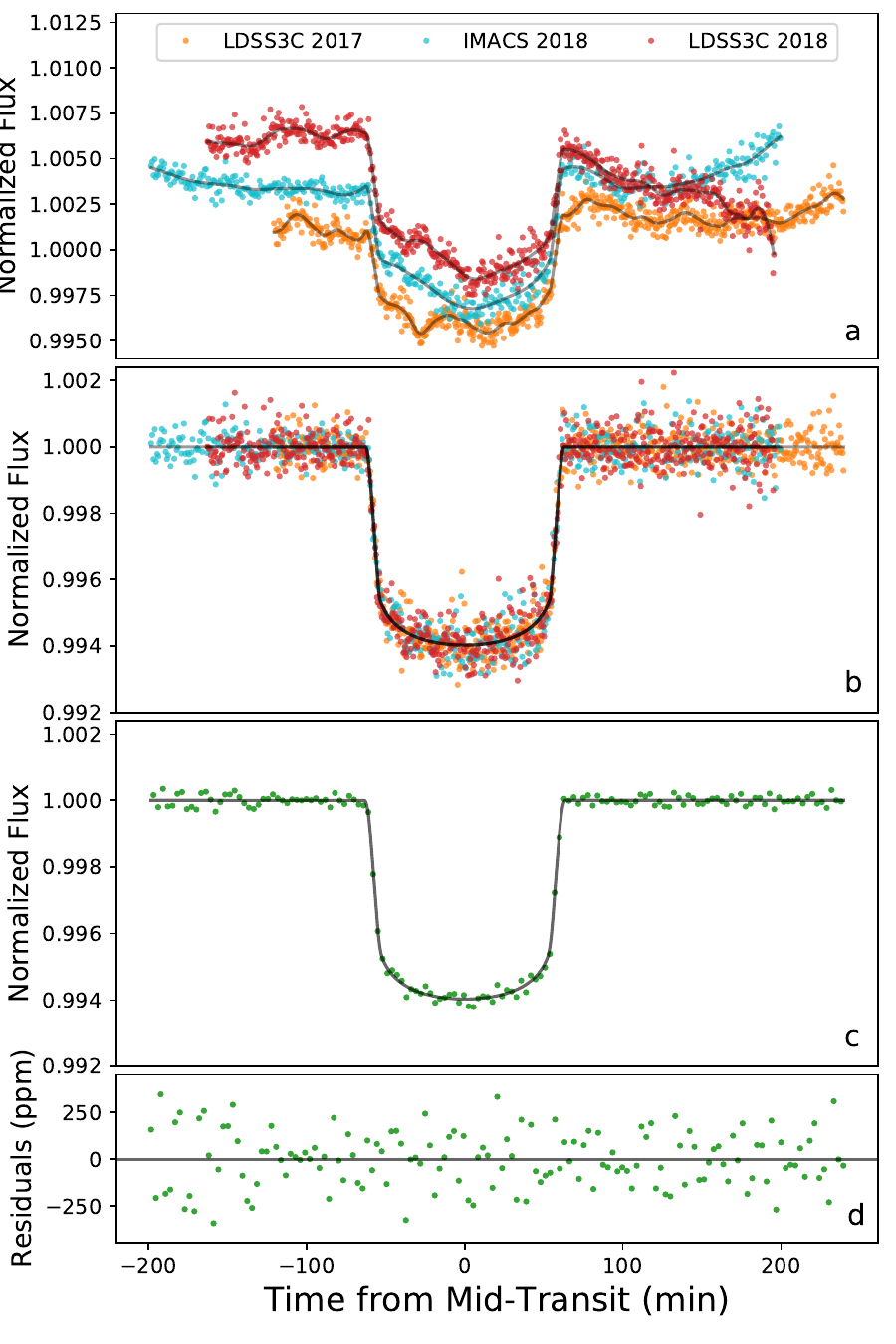}
\caption{(a) Raw white light curves from the three data sets used in this work. Overplotted gray lines are the fitted models to each data set using the GP regression method. (b) Light curves with the noise component of the model divided out, leaving just the transit data and model. (c) All of the data are combined and binned into 3-minute time bins. The transit model is sampled at high cadence and smoothed with a 3-minute boxcar kernel. (d) Residuals of panel (c). The rms of the 3-minute time-binned white light curve (panel (c)) is 145 ppm; binning down to 10 minutes gives an rms of 77 ppm. Data in all three data sets are summed from 610 to 950 nm, where the wavelength range in common to all three data sets.}
\label{fig:whitelightcurve}
\end{figure}

Once the input vectors associated with each data set are determined by analyzing them separately, we then use those input vectors to perform a joint fit across all three data sets. To construct this white light curve, we only use those data that correspond to wavelengths common to all data sets. In the case of the LHS 1140b white light curve we use data from 610 to 950 nm, where all three data sets have measured fluxes (as can be seen in Figure~\ref{fig:spectrabinned}. The raw, decorrelated, and time-binned white light curves are shown in Figure~\ref{fig:whitelightcurve}. We achieve an rms of 145 ppm when binning the fitted joint white light curves to 3-minute time bins and an rms of 77 ppm when binning to 10-minute time bins. The parameters we use or derive from the joint white light-curve fit are presented in Table~\ref{tab:whitelcvalues}, along with their priors.

We note that it is not technically correct to assume the same limb-darkening values for the LDSS3C and IMACS data sets, since these two instruments do not produce the same spectral energy distribution. However, this difference is not likely to affect the times of midtransit derived from the white light-curve fit, which is the only parameter we take from the white light-curve fit and use in the wavelength-binned fits.

The \citet{Ment2019} analysis included high-cadence (2 s integrations) Spitzer data. We adopt the period $P$, inclination $i$, and semi-major axis $a/R_s$ derived from that work as fixed parameters in the white-light and wavelength-binned light-curve analysis.

We then bin the light curves into 20 nm bins. We analyze the three data sets jointly in each wavelength bin, but the wavelength bins are independently analyzed from each other. We fix the orbital parameters that are common to all wavelength bins to their fitted values or literature values. In Figure 6 we compare independently fit wavelength-binned transit depths to those that are jointly fit. We also compare GP versus linear fitting methods. It is apparent that the GP fit produces larger error bars, as it should, but also more scatter in transit depths. This could be due to the fact that we used a minimal number of input vectors in the GP fit in order to reduce computation time.

\begin{figure*}
\includegraphics[width=1\textwidth]{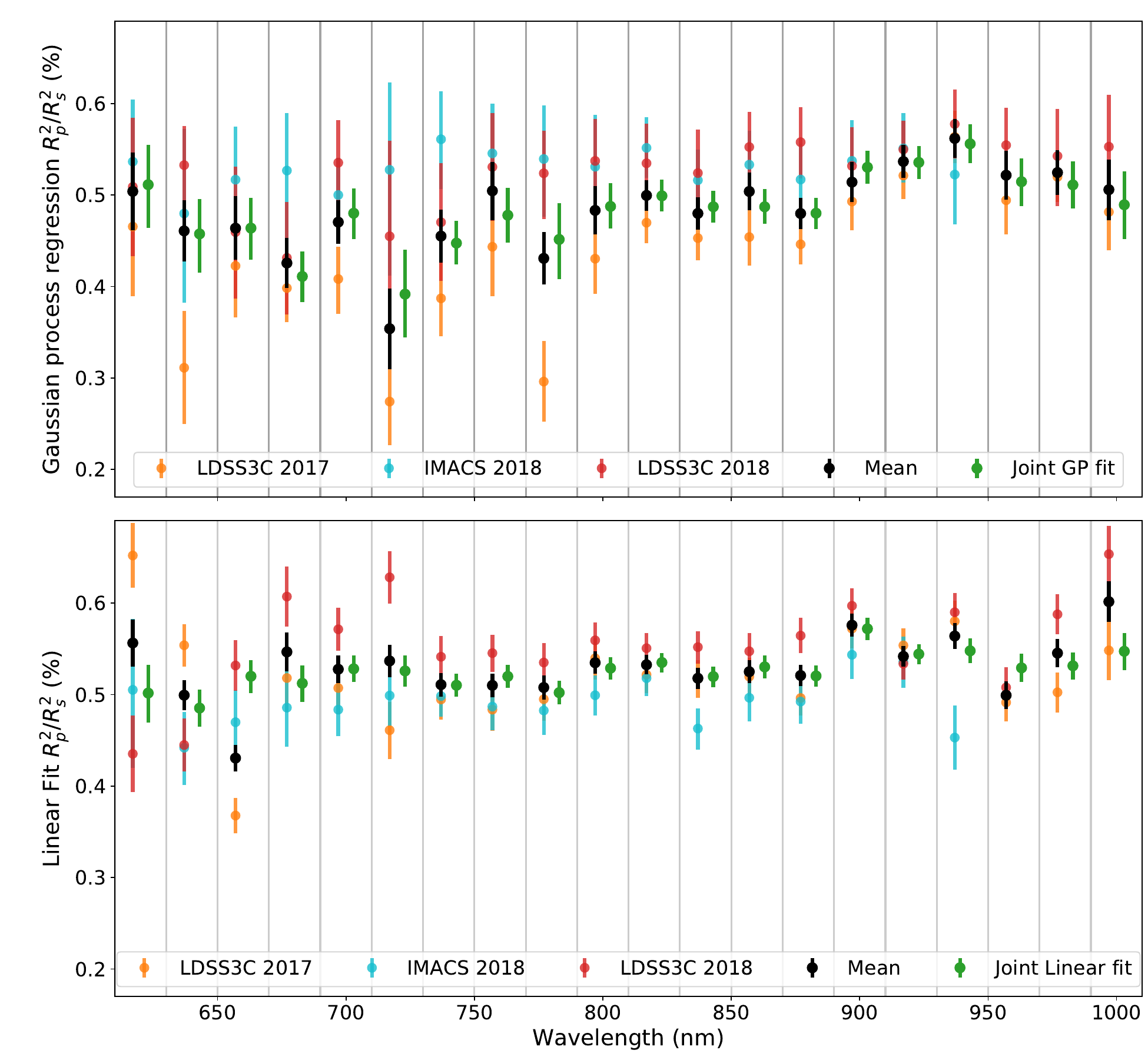}
\caption{Comparison of independent and jointly fit wavelength-binned transit depths, and of transit depths computed with a GP regression and a linear fit. Top: the independently fit transit depths from the LDSS3C 2017, IMACS 2018, and LDSS3C 2018 observations are presented, with the inverse-variance-weighted mean of these values overplotted in black. The jointly fit transit depths are presented in green. Error bars represent 1$\sigma$ uncertainties in the transit depth.  Vertical gray lines denote the limits of the wavelength bins, as in Figure~\ref{fig:spectrabinned}. The transit depths are offset in the x-axis for clarity; the analyses were performed over the same wavelength ranges in the independent and joint fits. Bottom: same as the top panel, but transit depths are derived using a linear fit. Transit depths for each transmission spectrum presented here are provided in Table~\ref{tab:transitdepths}.} 
\label{fig:comapretransmission}
\end{figure*}

\begin{figure}
\includegraphics[width=0.48\textwidth]{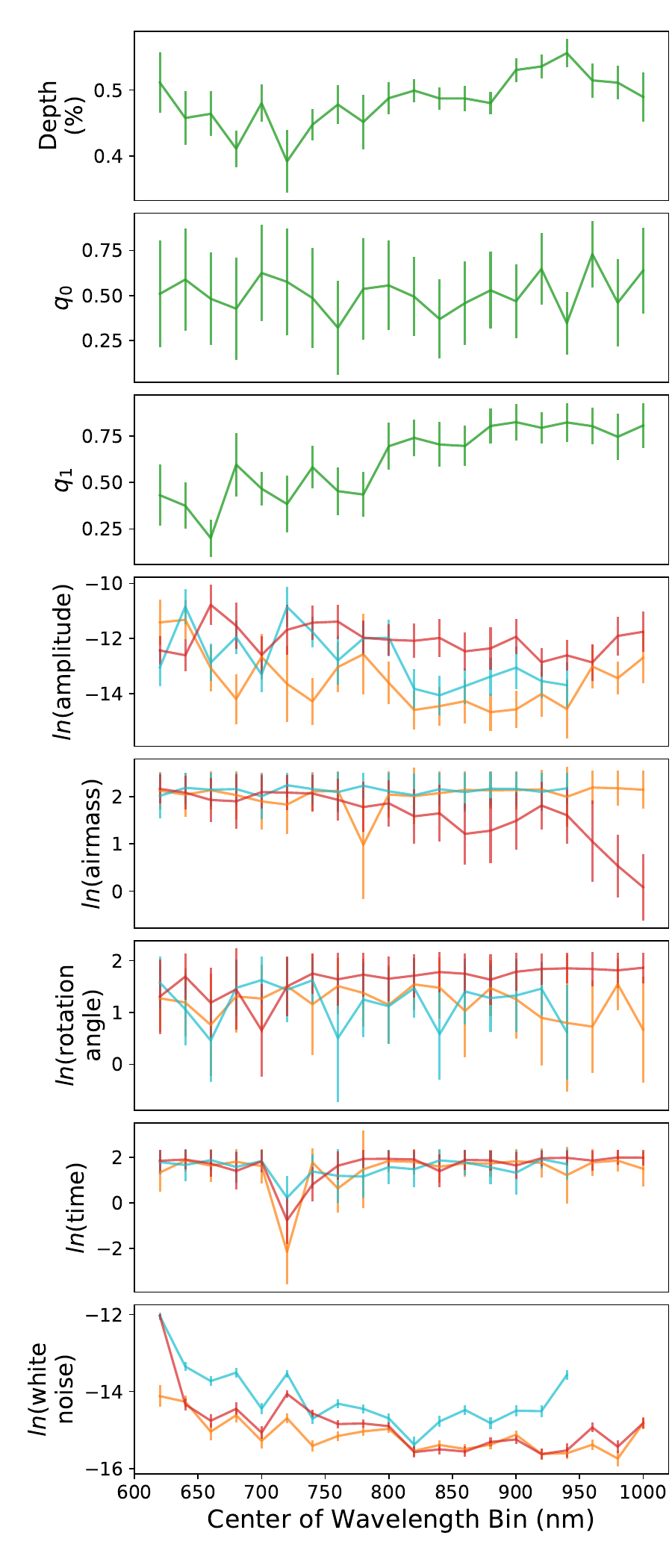}
\caption{Results from the wavelength-binned joint fit using a GP regression. The radius ratio and scaled limb-darkening coefficients are shared across all three data sets, and so there is only one resulting value in each wavelength bin (green points with error bars). The rest of the parameters are fit simultaneously, but separately for each data set (colors correspond to the same data sets as in Figure~\ref{fig:whitelightcurve}). We do not see any obvious correlations between any of the parameters and the resulting measurement of $R_p^2/R_s^2$.}
\label{fig:parameters}
\end{figure}

\begin{figure*}
\includegraphics[width=1\textwidth]{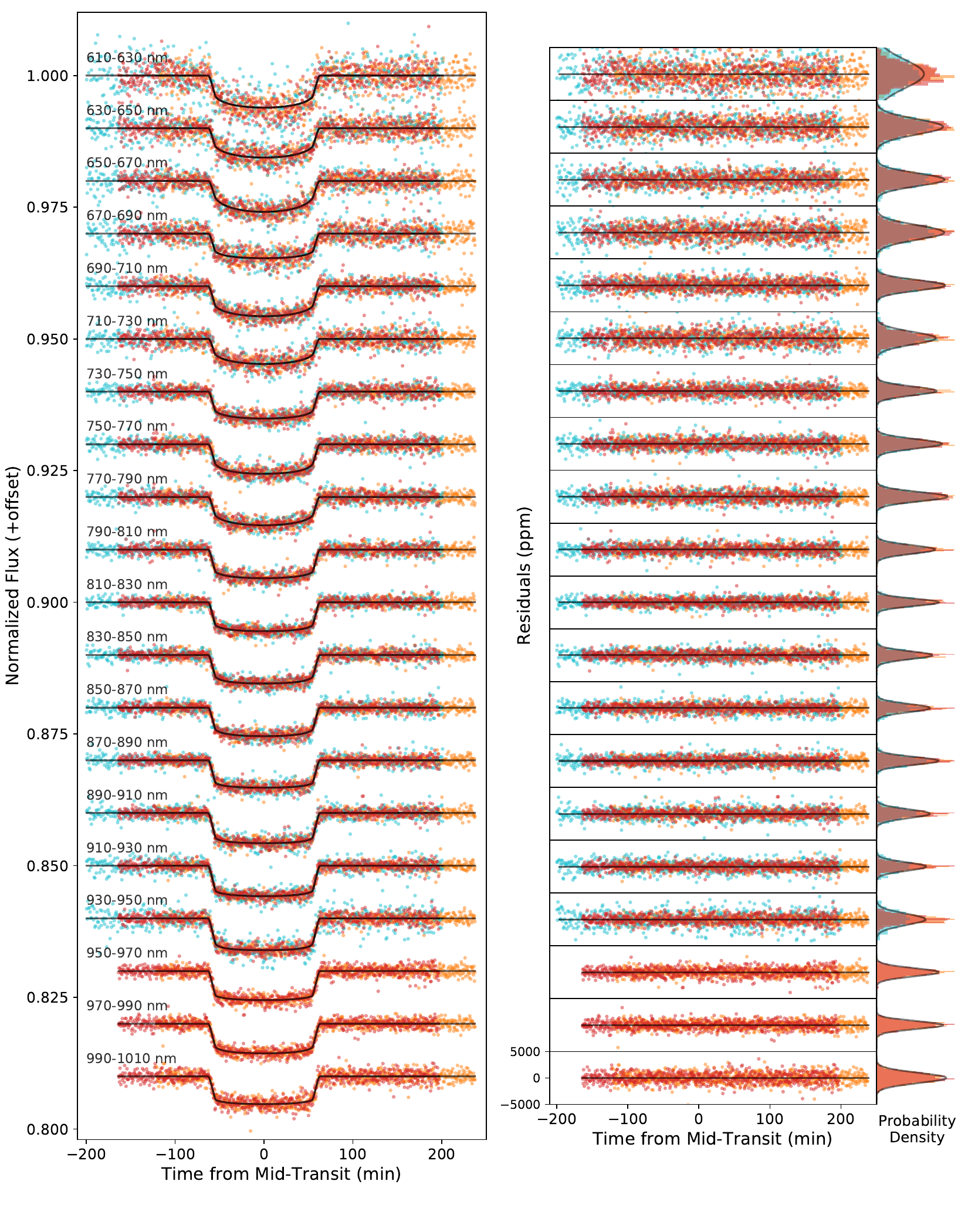}
\vspace{-35pt}
\caption{Left: light curves in each 20 nm wavelength bin with noise components modeled with a GP regression and removed. Right: residuals in each wavelength bin. The $y$-axes of each residual panel are the same, but for clarity we only label them in the bottom panel. We also show the residual histograms for each data set, compared to a Gaussian distribution (gray line) of the combined residuals from all three data sets. Because each of the three component data sets has a different Poisson noise, the three-component Gaussians are not expected to be identical, and hence the single gray curve is simply meant to be representative of the average behavior. The colors of the points and histograms in this figure represent the three data sets and follow the same color scheme as in Figure~\ref{fig:whitelightcurve}}
\label{fig:wavebinlightcurve}
\end{figure*}

In Figure~\ref{fig:parameters} we graphically present the final parameter values from the joint fit GP regression in each wavelength bin. We present the wavelength-binned data along with the best light-curve fits from the GP regression in Figure~\ref{fig:wavebinlightcurve}. Table~\ref{tab:transitdepths} provides the measured values of $R_p^2/R_s^2$ for the GP and linear fitting methods, as well as a further breakdown to the independently and jointly fit transit depths. We also provide the rms of each set of transit depths in the table, as compared to the inverse-variance-weighted mean of each transmission spectrum. At the end we provide light-curve rms for each wavelength bin in the GP joint fit, along with how close we were able to get to the photon noise limit in that bin. Across all 20 wavelength bins we achieve an average uncertainty in $R_p^2/R_s^2$ of 0.026\% (260 ppm) and an average rms value of 1.3$\times$ the photon noise. 

\movetabledown=7in
\begin{rotatetable*}
\begin{threeparttable}
\caption{Best-fit $R_p^2/R_s^2$ for Linear and Gaussian Process Regressions \label{tab:transitdepths}}
\definecolor{Gray}{gray}{0.85}
\newcolumntype{g}{>{\columncolor{Gray}[1.65mm][1.65mm]}c}
\begin{tabular}{cggggcccccc} 
\hline
\hline
Wavelength & \multicolumn{4}{c}{Linear Fit Transit Depths (\%)} & \multicolumn{4}{c}{GP Fit Transit Depths (\%)} & rms    & $\times$ Expected \\
(nm)       & L17 & I18 & L18 & Joint & L17 & I18 & L18 & Joint    & (ppm)  &  Noise            \\
\hline
610-630 & 0.652 $\pm$ 0.035 & 0.505 $\pm$ 0.081 & 0.435 $\pm$ 0.042 & 0.502 $\pm$ 0.032 & 0.466 $\pm$ 0.073 & 0.536 $\pm$ 0.072 & 0.509 $\pm$ 0.076 & 0.511 $\pm$ 0.511 & 2659 & 1.51 \\ 
630-650 & 0.554 $\pm$ 0.023 & 0.442 $\pm$ 0.040 & 0.445 $\pm$ 0.029 & 0.485 $\pm$ 0.020 & 0.311 $\pm$ 0.062 & 0.480 $\pm$ 0.095 & 0.533 $\pm$ 0.044 & 0.458 $\pm$ 0.458 & 1464 & 1.27 \\ 
650-670 & 0.368 $\pm$ 0.020 & 0.470 $\pm$ 0.035 & 0.532 $\pm$ 0.028 & 0.520 $\pm$ 0.018 & 0.423 $\pm$ 0.054 & 0.517 $\pm$ 0.059 & 0.459 $\pm$ 0.072 & 0.464 $\pm$ 0.464 & 1217 & 1.23 \\ 
670-690 & 0.518 $\pm$ 0.037 & 0.486 $\pm$ 0.042 & 0.607 $\pm$ 0.033 & 0.512 $\pm$ 0.020 & 0.398 $\pm$ 0.034 & 0.527 $\pm$ 0.069 & 0.431 $\pm$ 0.062 & 0.411 $\pm$ 0.411 & 1376 & 1.25 \\ 
690-710 & 0.507 $\pm$ 0.028 & 0.483 $\pm$ 0.029 & 0.571 $\pm$ 0.023 & 0.528 $\pm$ 0.015 & 0.408 $\pm$ 0.037 & 0.500 $\pm$ 0.042 & 0.535 $\pm$ 0.048 & 0.480 $\pm$ 0.480 & 989 & 1.20 \\ 
710-730 & 0.461 $\pm$ 0.031 & 0.499 $\pm$ 0.033 & 0.628 $\pm$ 0.029 & 0.526 $\pm$ 0.017 & 0.274 $\pm$ 0.056 & 0.528 $\pm$ 0.106 & 0.455 $\pm$ 0.100 & 0.392 $\pm$ 0.392 & 1210 & 1.40 \\ 
730-750 & 0.495 $\pm$ 0.022 & 0.498 $\pm$ 0.022 & 0.541 $\pm$ 0.023 & 0.510 $\pm$ 0.013 & 0.387 $\pm$ 0.041 & 0.561 $\pm$ 0.053 & 0.470 $\pm$ 0.065 & 0.447 $\pm$ 0.447 & 849 & 1.36 \\ 
750-770 & 0.483 $\pm$ 0.023 & 0.487 $\pm$ 0.025 & 0.545 $\pm$ 0.020 & 0.520 $\pm$ 0.013 & 0.443 $\pm$ 0.053 & 0.546 $\pm$ 0.054 & 0.531 $\pm$ 0.059 & 0.478 $\pm$ 0.478 & 894 & 1.36 \\ 
770-790 & 0.495 $\pm$ 0.023 & 0.483 $\pm$ 0.026 & 0.535 $\pm$ 0.021 & 0.502 $\pm$ 0.013 & 0.296 $\pm$ 0.044 & 0.539 $\pm$ 0.061 & 0.524 $\pm$ 0.048 & 0.451 $\pm$ 0.451 & 887 & 1.36 \\ 
790-810 & 0.540 $\pm$ 0.023 & 0.499 $\pm$ 0.022 & 0.559 $\pm$ 0.020 & 0.529 $\pm$ 0.012 & 0.430 $\pm$ 0.038 & 0.531 $\pm$ 0.060 & 0.537 $\pm$ 0.047 & 0.488 $\pm$ 0.488 & 823 & 1.40 \\ 
810-830 & 0.522 $\pm$ 0.020 & 0.518 $\pm$ 0.020 & 0.551 $\pm$ 0.017 & 0.535 $\pm$ 0.010 & 0.470 $\pm$ 0.021 & 0.552 $\pm$ 0.033 & 0.535 $\pm$ 0.044 & 0.499 $\pm$ 0.499 & 710 & 1.23 \\ 
830-850 & 0.517 $\pm$ 0.021 & 0.463 $\pm$ 0.022 & 0.552 $\pm$ 0.018 & 0.520 $\pm$ 0.011 & 0.453 $\pm$ 0.023 & 0.516 $\pm$ 0.034 & 0.524 $\pm$ 0.047 & 0.487 $\pm$ 0.487 & 760 & 1.29 \\ 
850-870 & 0.520 $\pm$ 0.021 & 0.496 $\pm$ 0.026 & 0.547 $\pm$ 0.020 & 0.530 $\pm$ 0.012 & 0.454 $\pm$ 0.031 & 0.533 $\pm$ 0.037 & 0.552 $\pm$ 0.039 & 0.487 $\pm$ 0.487 & 783 & 1.30 \\ 
870-890 & 0.496 $\pm$ 0.019 & 0.492 $\pm$ 0.024 & 0.564 $\pm$ 0.019 & 0.520 $\pm$ 0.012 & 0.446 $\pm$ 0.021 & 0.517 $\pm$ 0.039 & 0.558 $\pm$ 0.039 & 0.480 $\pm$ 0.480 & 771 & 1.31 \\ 
890-910 & 0.572 $\pm$ 0.022 & 0.543 $\pm$ 0.026 & 0.597 $\pm$ 0.019 & 0.572 $\pm$ 0.012 & 0.493 $\pm$ 0.031 & 0.537 $\pm$ 0.045 & 0.532 $\pm$ 0.042 & 0.530 $\pm$ 0.530 & 836 & 1.35 \\ 
910-930 & 0.554 $\pm$ 0.019 & 0.536 $\pm$ 0.028 & 0.534 $\pm$ 0.017 & 0.544 $\pm$ 0.011 & 0.521 $\pm$ 0.026 & 0.551 $\pm$ 0.038 & 0.550 $\pm$ 0.032 & 0.536 $\pm$ 0.536 & 808 & 1.26 \\ 
930-950 & 0.580 $\pm$ 0.022 & 0.453 $\pm$ 0.035 & 0.590 $\pm$ 0.021 & 0.548 $\pm$ 0.014 & 0.564 $\pm$ 0.029 & 0.522 $\pm$ 0.055 & 0.578 $\pm$ 0.037 & 0.556 $\pm$ 0.556 & 1089 & 1.24 \\ 
950-970 & 0.491 $\pm$ 0.021 & --- & 0.508 $\pm$ 0.022 & 0.529 $\pm$ 0.016 & 0.494 $\pm$ 0.036 & 0.522 $\pm$ 0.055 & 0.554 $\pm$ 0.040 & 0.515 $\pm$ 0.515 & 789 & 1.30 \\ 
970-990 & 0.502 $\pm$ 0.022 & --- & 0.588 $\pm$ 0.022 & 0.531 $\pm$ 0.015 & 0.520 $\pm$ 0.027 & 0.522 $\pm$ 0.055 & 0.543 $\pm$ 0.053 & 0.511 $\pm$ 0.511 & 742 & 1.18 \\ 
990-1010 & 0.548 $\pm$ 0.032 & --- & 0.653 $\pm$ 0.031 & 0.547 $\pm$ 0.020 & 0.481 $\pm$ 0.041 & 0.522 $\pm$ 0.055 & 0.553 $\pm$ 0.057 & 0.489 $\pm$ 0.489 & 972 & 1.24 \\ 
\hline
rms (ppm) & 540 & 257 & 513 & 186 & 736 & 193 & 380 & 388 & &\\ 
\hline
\end{tabular}
\begin{tablenotes}
\item[1] \textbf{Note.} In this table we present transit depths and uncertainties derived from both a linear and GP regression. We also present the transit depth of each data set individually, along with the joint fits, in which the transit depth was a shared parameter between the three data sets. These values correspond to those in Figure~\ref{fig:comapretransmission}. The last row provides the rms of each column of transit depths, as compared to the inverse-variance-weighted mean (i.e., a flat line) of that column. The final two columns pertain to the joint GP fit (third-to-last column) and provide the light-curve rms for each wavelength bin and compare this to the expected noise for that wavelength bin.
\end{tablenotes}
\end{threeparttable}
\end{rotatetable*}

\section{Results and Discussion} \label{sec:results}

From our observations we produce a transmission spectrum and compare it to models. In doing so, we demonstrate the limits of the ground-based transmission spectroscopy technique employed here to investigate the atmosphere of LHS 1140b.

\subsection{Planetary atmospheric detection}\label{subsec:atmodetection}
For the purposes of transmission spectroscopy, we are interested in the scale height $H$ of a planet's atmosphere, or how extended the atmosphere is, and what kinds of features it produces. The scale height is calculated by

\begin{equation}\label{eqn:H}
    H = \frac{k_BT}{\mu g}
\end{equation}

\noindent where $k_B$ is the Boltzmann constant, $T$ is the planet's mean atmospheric temperature, $\mu$ is the mean molecular weight of the planet's atmosphere, and $g$ is the planet's surface gravity. We do not know the mean temperature of LHS 1140b's atmosphere, but because the transmission spectrum is an integration of light paths across multiple atmospheric layers, it is not sensitive to varying temperature gradients throughout the atmosphere \citep{Kempton2017}. We can therefore estimate a temperature--pressure profile for the transmission spectrum using the planet's equilibrium temperature.

An estimate of the amplitude of features in the transmission spectrum of an atmosphere is given by 

\begin{equation}\label{eqn:deltad}
    \begin{aligned}
        \Delta \delta &= \left(\frac{R_p + NH}{R_s}\right)^2 - \left(\frac{R_p}{R_s}\right)^2\\
        &\approx \frac{2R_pNH}{R_s^2}
    \end{aligned}
\end{equation}

\noindent where $N$ is the number of scale heights we can observe before the atmosphere becomes optically thick (when optical depth $\tau\gg1$).

\subsection{Model transmission spectrum}\label{subsec:exotransmit}

The relative feature amplitudes of a planetary atmosphere observed over a range of wavelengths can be compared to models in order to reveal the presence of an atmosphere and its composition. We construct a model transmission spectrum for LHS 1140b using the open-source code \texttt{Exo-Transmit} \citep{Miller-RicciKempton2012,Kempton2017}. The code inputs are a temperature--pressure profile, an equation of state specific to the atmospheric composition, the 1-bar planet radius and surface gravity, and the stellar radius. 

Following the same procedures outlined in \cite{Miller-Ricci2009} and \citet{Miller-Ricci2010}, we use custom double-gray temperature--pressure profiles for the LHS 1140b atmosphere. (The default temperature--pressure profiles that comes with \texttt{Exo-Transmit} are isothermal.) The equation-of-state files corresponding to the atmospheres we test in this work are readily available in \texttt{Exo-Transmit}. Since we do not know the 1-bar planet radius exactly, we adjust it until the model transmission atmosphere best fits the data. This adjustment changes both the absolute depth of the model and the amplitude of the features.

\subsection{Observed transmission spectrum}\label{subsec:transmission}

\begin{figure*}
\includegraphics[width=1\textwidth]{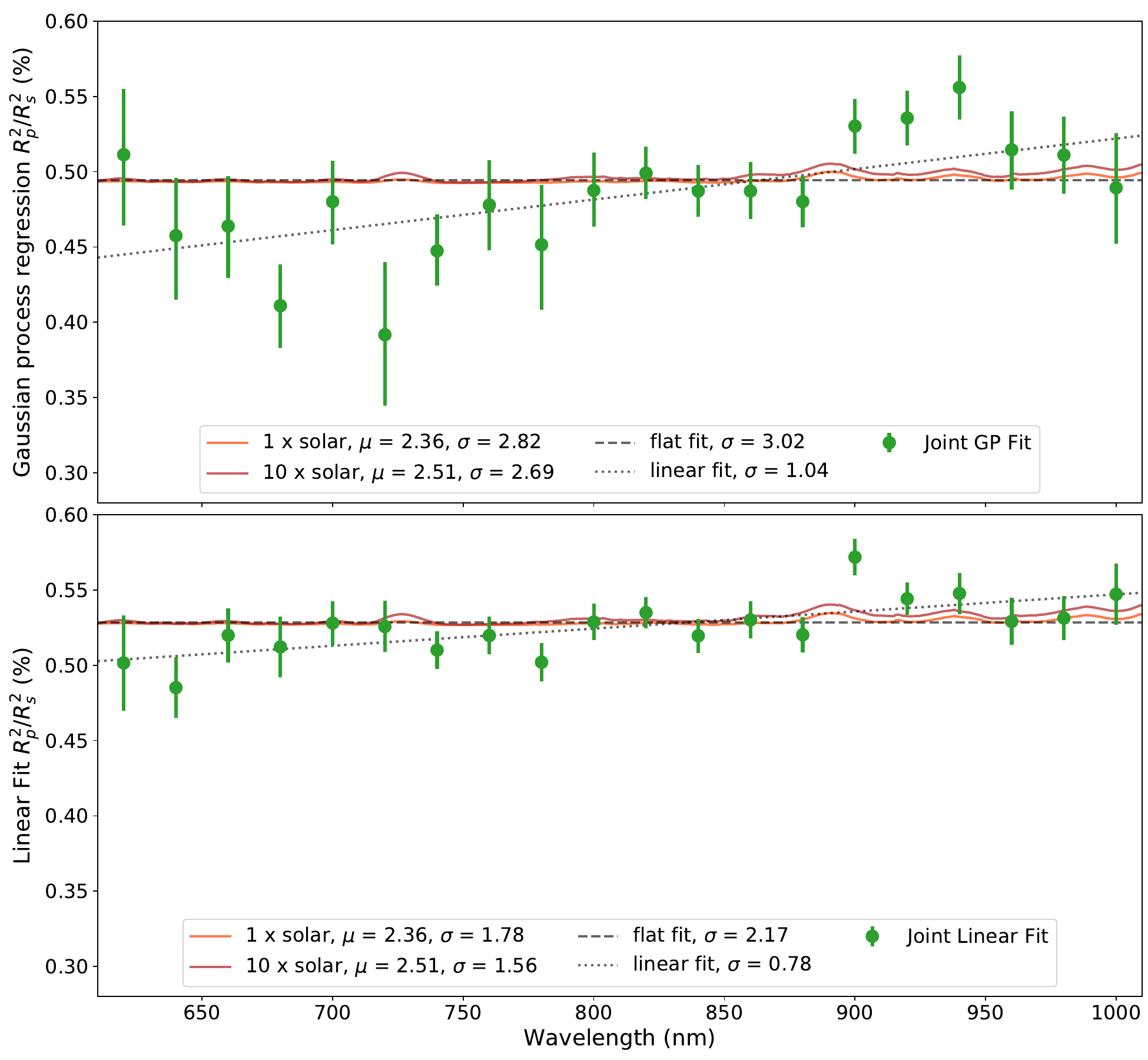}
\caption{Transmission spectrum of LHS 1140b. Top transmission spectrum constructed using a GP regression to jointly fit three data sets taken with the IMACS and LDSS3C spectrographs on Magellan I (Baade) and II (Clay), respectively. We compare the observed transmission spectrum (green points with 1$\sigma$ error bars) to model transmission spectra with compositions that are 1$\times$ and 10$\times$ solar metallicity by volume (orange and red lines, respectively). For the model transmission spectra we state the mean molecular weight associated with each model in the legend. The gray dashed line is the inverse-variance-weighted mean of the observed transmission spectrum, while the gray dotted line is a linear fit. We also state the $\sigma$-confidence with which we rule out the fit or model presented. The linear fit is marginally better than the flat fit; however, none of the models or fits are significantly disfavored to the precision that we report in the figure, so none can be ruled out by the observations. Bottom: same as the top panel, but using a linear regression to jointly fit the data sets.}
\label{fig:transmission}
\end{figure*}

From the wavelength-binned jointly fitted $R_p/R_s$ values, we construct a transmission spectrum. In Figure~\ref{fig:transmission} we present the final transmission spectrum and compare it to model transmission spectra calculated for the \mbox{LHS 1140b} system using \texttt{Exo-Transmit} \citep{Kempton2017}.

Given the small atmospheric features of the LHS 1140b atmosphere, we are not able to rule out even the lowest mean molecular weight cases. We therefore only present these cases -- clear 1$\times$ and 10$\times$ solar metallicity atmospheres -- and do not address models of higher mean molecular weight atmospheres. We do not expect a terrestrial planet like LHS 1140b to possess such light atmospheres \citep{Lopez2013,Owen&Wu2013,Rogers2015,Fulton2017,VanEylen2018}, but these end-member compositions are the first atmospheres to rule out.

We note that the points in the range of 890-950 nm appear systematically high. Indeed, removing these three points from the GP joint fit transmission spectrum yields $\chi^2=18$ for the flat line fit, as opposed to $\chi^2=41$ when the three points are included. Between 890 and 950 nm is a forest of water lines. If there is water present in starspots on the photosphere of LHS 1140, which is possible given its effective temperature of $3216 \pm 39$ \citep{Ment2019}, then this could artificially deepen the transit depths in the water band if there are unocculted inhomogeneities on the stellar surface during observations. Alternatively, telluric water may persist in the transmission spectrum, despite our use of a comparison star to remove such telluric features; recall that we must use a comparison star of a different spectral type, which would experience different rates of extinction in the presence of additional water in the air column through which we observe. 

\subsection{Atmospheric detection limits}

To explore the limits of our observed transmission spectrum, we can perform simple calculations using Equations~\ref{eqn:H} and~\ref{eqn:deltad}. LHS 1140b's surface gravity ($23.7 \pm 2.7$ m/s$^2$) and cool equilibrium temperature \citep[$235 \pm 5$ K, assuming a Bond albedo of 0 and a planet-wide energy distribution;][]{Ment2019} combine to make the scale height of this planet's atmosphere $40.9 \pm 4.7$ km for the lowest mean molecular weight case ($\mu = 2$) for the unrealistic, pure light H$_2$ atmosphere.

For the LHS 1140 system where $R_p = 1.727 \pm 0.032\ R_{\oplus}$ and $R_s = 0.2139 \pm 0.0014\ R_{\odot}$, the amplitude of the transmission features for the lowest mean molecular weight case is $65.3 \pm 7.7$ ppm, assuming that we can see down 1.6 scale heights. (We estimate $N = 1.6$ from the 20 nm wavelength-binned model transmission spectrum.) This is a factor of four below the median precision we are able to achieve in this project. For a more realistic atmosphere dominated by CH$_4$, H$_2$O, O$_2$, or CO$_2$, the feature amplitudes are at the level of 8 ppm or lower, a factor of 35 below our precision.

\subsection{Future instruments}

The heavily anticipated James Webb Space Telescope (JWST) will be capable of robust detections of planetary atmospheres with instruments capable of performing transmission spectroscopy across a broad wavelength range. \citet{Morley2017} simulated JWST observations with NIRSpec/G235M and the F170LP filter for several nearby planets orbiting small stars, assuming equilibrated atmospheres derived from Titan, Earth, and Venus elemental compositions. The authors conclude that it will not be possible to detect an atmosphere around LHS 1140b with JWST owing to an unrealistic amount of observing time. This conclusion is still valid, despite a refinement of some of the LHS 1140 system parameters \citep{Ment2019}.

The next generation of ground-based optical telescopes---the Giant Magellan Telescope, the Thirty Meter Telescope, and the European Extremely Large Telescope---will be larger than any we currently have. All three have planned multiobject spectrographs as either first-light or second-generation instruments. Exposure time calculators are not yet available for these modes, but a simple scaling to the larger collecting areas of these telescopes reveals that the low mean molecular weight atmospheres tested in this study could be ruled out on LHS 1140b with as few as seven transits with GMT or five transits with TMT. These ground-based observatories will still have to contend with LHS 1140b's infrequent transits. High-resolution spectroscopy \citep{Snellen2013,Birkby2018} will be possible with the GSMTs, but LHS 1140b will still likely be below the detection thresholds for this technique.

Perhaps the most promising avenue for detecting the atmospheres of habitable-zone terrestrial exoplanets is to find more amenable targets. As TESS continues to discover new worlds around our closest stellar neighbors, we are likely to find planets with more accessible atmospheres than that of LHS 1140b. TESS has already grown the sample of nearby ($<15$ pc) terrestrial exoplanets, including prime targets such as LHS 3844b \citep{Vanderspek2019} and LTT 1445Ab \citep{Winters2019}, though neither planet resides in the habitable zone.

\section{Conclusion}\label{sec:conclusion}

LHS 1140b orbits in the habitable zone of its host M dwarf. This world is at the upper end of the radius regime that defines terrestrial planets \citep{Fulton2017}, but we know from radius and mass measurements that it is rocky in nature \citep{Ment2019}. However, given the high surface gravity and cool equilibrium temperature of LHS 1140b, its atmosphere is not readily accessible to transmission spectroscopy. 

With this work we set out to capture two transits of LHS 1140b while also exploring the synergy between the IMACS and LDSS3C spectrographs. Because LHS 1140b transits infrequently, ground-based opportunities for observation are rare. We designed a multiyear program that employed both Magellan I/IMACS and II/LDSS3C, though LDSS3C is preferred for M dwarf observations because its red observing mode collects more than twice as many photons as IMACS at the wavelengths where M dwarfs emit the bulk of their photons (Figure~\ref{fig:spectrabinned}).

Though we are not able to investigate the atmosphere of LHS 1140b in this work, we detail our extraction and analysis pipelines in order to illustrate how we convert raw spectroscopic information into wavelength-calibrated time series. We construct both a white light curve and 20 nm wavelength-binned light curves by jointly fitting our data sets. In the joint fit white light curves, we achieve an rms of 145 ppm when binning the data to 3-minute time bins; we achieve an rms of 77 ppm when binning to 10-minute time bins. The data bin down predictably with the calculated rms. Large ground-based telescopes like the Magellans can be used as high-precision white-light photometers to probe deviations in transit shape that may arise from, for example, planetary oblateness or moons.

Across all of the wavelength-binned light curves we achieve an average uncertainty in $R_p^2/R_s^2$ of 0.028\% and an average precision in the wavelength-binned light curves of 1.3$\times$ the photon noise. We will employ the techniques laid out in this work for ground-based transmission spectroscopy studies of the recently discovered terrestrial worlds LHS 3844b and LTT 1445Ab \citep{Vanderspek2019,Winters2019} with Magellan II (Clay)/LDSS3C (PI Diamond-Lowe). These worlds do not reside in the habitable zones of their systems, but they are more amenable to atmospheric follow-up.

Finally, in the TESS era, we emphasize the need for robust mass and radius measurements of newly discovered transiting exoplanets. Without knowledge of the bulk densities of these worlds we will under- or overestimate our ability to detect their atmospheres. 

\acknowledgments
We thank the anonymous referee for their detailed comments, which greatly enhanced the scope of this work in terms of data analysis methods. This paper includes data gathered with both of the 6.5m Magellan Telescopes (Baade and Clay) located at Las Campanas Observatory, Chile. We thank the contributors to the IMACS and LDSS3C projects, the telescope operators and staff at Las Campanas Observatory, and the writers and contributors of the open-source software used in this work. We make use of the Digitized Sky Surveys in Figure 1 of this paper, as well as during observations. The Digitized Sky Surveys were produced at the Space Telescope Science Institute under U.S. Government grant NAG W-2166. The images of these surveys are based on photographic data obtained using the Oschin Schmidt Telescope on Palomar Mountain and the UK Schmidt Telescope. The plates were processed into the present compressed digital form with the permission of these institutions. We especially thank members of the ACCESS Collaboration N\'estor Espinoza, Benjamin Rackham, David Osip, and Mercedes Lopez-Morales for in-depth conversations about the workings of the IMACS multiobject spectrograph. We also thank Robin Wordsworth, Dimitar Sasselov, Laura Kreidberg, James Kirk, and Amber Medina for helpful comments and conversations. We thank Erik Strand for assistance during the 2017 observations. H.D.-L. recognizes support from the National Science Foundation Graduate Research Fellowship Program (grant No. DGE1144152). J.A.D. would like to acknowledge support from the Heising-Simons Foundation for their support, whose 51 Pegasi b Postdoctoral Fellowship program has enabled this work. The work of E.M.-R.K. was supported by the National Science Foundation under grant No. 1654295 and by the Research Corporation for Science Advancement through their Cottrell Scholar program. This publication was made possible through the support of a grant from the John Templeton Foundation. The opinions expressed here are those of the authors and do not necessarily reflect the views of the John Templeton Foundation.

\vspace{5mm}
\facilities{Magellan I Baade (IMACS), Magellan II Clay (LDSS3C)}

\software{\texttt{astropy} \citep{AstropyCollaboration2013,AstropyCollaboration2018}, \texttt{batman} \citep{Kreidberg2015}, \texttt{decorrasaurus} (\href{https://github.com/hdiamondlowe/decorrasaurus/releases/tag/v2.0}{github.com/hdiamondlowe/decorrasaurus}), \texttt{dill} \citep{McKerns2012}, \texttt{dynesty} \citep{Speagle2020}, \texttt{Exo-Transmit} \citep{Kempton2017}, \texttt{george} \citep{Foreman-Mackey2015}, \texttt{LDTk} \citep{Parviainen2015}, \texttt{mosasaurus} (\href{http://www.github.com/zkbt/mosasaurus}{github.com/ zkbt/mosasaurus}), \texttt{SAOImageDS9} \citep{Joye2003}}

\bibliography{MasterBibliography.bib}

\end{document}